\documentclass{aa}  
\usepackage{hyperref}
\usepackage{graphicx}
\usepackage{txfonts}
\usepackage{natbib}
\usepackage{mathrsfs,mathpazo,amsmath,mathtools,physics}
\usepackage{dcolumn}
\usepackage{hyperref}
\usepackage{xcolor}
\usepackage{ulem}
\usepackage{tikz}
\usepackage{listings}
\usepackage{lipsum, babel}
\usepackage{multirow}
\usepackage{float}
 
\usepackage[font=small,labelfont=bf,justification=justified]{subcaption}
\usepackage[font=small,labelfont=bf,justification=justified]{caption}

\begin{document}

\newcommand{\fmc}{\;fm$^{-3}$\;}
\newcommand{\gmc}{\;g/cm$^{-3}$\;}
\newcommand{\mdot}{\;M$_{\odot}$\;}
\newcommand{\nsat}{$n_{\rm sat}$\;}
\newcommand{\esat}{$E_{\rm sat}$\;}
\newcommand{\ksat}{$K_{\rm sat}$\;}
\newcommand{\qsat}{$Q_{\rm sat}$\;}
\newcommand{\zsat}{$Z_{\rm sat}$\;}
\newcommand{\esym}{$E_{\rm sym}$\;}
\newcommand{\lsym}{$L_{\rm sym}$\;}
\newcommand{\ksym}{$K_{\rm sym}$\;}
\newcommand{\qsym}{$Q_{\rm sym}$\;}
\newcommand{\zsym}{$Z_{\rm sym}$\;}
\newcommand{\expa}{SA-Exp-n$_0$\;}
\newcommand{\expb}{SA-Exp-2n$_0$\;}
\newcommand{\chiset}{SA-$\chi$-n$_0$\;}
\newcommand{\mmset}{MM-$\chi$\;}

\renewcommand{\prc}{PRC}
\renewcommand{\prd}{PRD}
\renewcommand{\prl}{PRL}
\renewcommand{\aap}{A\&A}
\renewcommand{\apj}{ApJ}
\renewcommand{\apjl}{ApJL}
\renewcommand{\mnras}{MNRAS}

\title{Nuclear parameter inference with semi-agnostic priors}

\author{Lami Suleiman \inst{1, 2}, Anthea F. Fantina\inst{3}, Francesca Gulminelli\inst{4}, Jocelyn Read\inst{5}}

\institute{Deutsches Elektronen-Synchrotron DESY, Platanenallee 6, 15738 Zeuthen, Germany \\
\email{lami.suleiman@desy.de} 
\and Deutsches Zentrum f\"ur Astrophysik (DZA), Postplatz 1, 02826 G\"orlitz, Germany
\and Grand Acc\'el\'erateur National d'Ions Lourds (GANIL), CEA/DRF - CNRS/IN2P3, Boulevard Henri Becquerel, 14076 Caen, France
\and Université de Caen Normandie, ENSICAEN, CNRS/IN2P3, LPC Caen UMR6534, F-14000 Caen and Institut Universitaire de France (IUF), France
\and Nicholas and Lee Begovich Center for Gravitational Wave Physics and Astronomy, California State University Fullerton, Fullerton, California 92831, USA.
}

\titlerunning{Nuclear parameter inference with semi-agnostic priors}
\authorrunning{Suleiman et al.}

\abstract
   {Radio pulsar timing, X-ray pulse profile modeling, and gravitational-wave detections of binary mergers involving at least one neutron star probe the properties of dense, neutron-rich matter in thermodynamic regimes inaccessible to nuclear laboratories. Such inference relies on building appropriate equation-of-state priors, such as the recently introduced semi-agnostic constructions that incorporate nuclear theory and experimental information available in low- to intermediate-density regimes, while offering the necessary flexibility at high density.}
     {In this paper, we assess how detections of mass, radius, and tidal deformability for low-mass ($\sim 1 M_\odot$) or high-mass ($\sim 1.9 M_\odot$) neutron stars contribute to constraining nuclear empirical parameters in an inference based on semi-agnostic equation-of-state priors.}
   {We first assessed the correlation factors between nuclear empirical parameters and the zero-temperature and $\beta$-equilibrated pressure in different density regimes. We then simulated observations for three nucleonic equations of state to test the recovery of the corresponding  nuclear empirical parameters.}
   {We show that not all nuclear empirical parameters significantly correlate with the pressure and find that they compete in the high-density regime, which challenges their inference. We also find that using semi-agnostic constructions instead of assuming a nucleonic content up to the highest densities in the neutron-star core can help recover the true nuclear empirical parameters with more accuracy.}
   {Parametrizing the high-density regime of the equation of state with the nucleonic meta-model can bias the inference of nuclear empirical parameters; semi-agnostic constructions provide a solution to this problem. However, many nuclear matter empirical parameters contribute similarly to the construction of the baryonic pressure. We find that they are difficult to infer independently, even with extremely precise measurements. 
   } 


\maketitle


\section{Introduction}\label{sec:intro}

Observations of neutron stars (NSs) can be used to explore the properties of dense matter under extreme conditions. The deepest layers of these compact objects are expected to reach densities that exceed several times the nuclear saturation density, well beyond the thermodynamic regimes explored by nuclear theory and experiments on Earth. 
Astrophysical features of NSs depend on the internal matter properties of the star. By comparing NS-modeled macroscopic parameters with measurements extracted from multi-messenger observations, 
we can constrain the properties of the strong interaction at relatively low temperatures and high densities.

A variety of signals emitted by NSs have already provided valuable information on their internal structure. The timing of radio pulsations emitted by the magnetic field of rotating NSs has enabled very accurate measurements of their mass (see e.g., \citet{Ozel2016}). The existence of high-mass pulsars such as PSR J0740$+$6620 \citep{Fonseca2021} requires that the pressure provided by matter in the high-density core of the star be sufficiently large to support the star against self-gravitational collapse. Gravitational waves (GWs) emitted during the merger of two compact objects retain the imprint of the masses and the deformation of the binary objects in the presence of an external gravitational field \citep{Dietrich2021}. The detection of the binary neutron star (BNS) merger GW170817 \citep{Abbott2017} indicates a preference for a small induced quadrupolar moment in response to an external tidal field, thus excluding high pressures in the NS core (see \citet{Abbott2018,Essick:2019ldf, Chatziioannou:2020pqz, Dietrich:2020efo, Wouters2025} and references therein). The relativistic effects revealed by the X-ray signal emitted by hot spots at the surface of millisecond pulsars allow us to measure the mass and radius of NSs simultaneously \citep{Watts2016}. The Neutron Star Interior Composition ExploRer (NICER) has published mass-radius posterior contours as well as equation of state (EoS) constraints for several sources (PSR J0030$+$0451 \citep{riley2019, miller2019}, PSR J0740$+$6620 \citep{riley2021, miller2021}, PSR J0437$-$4715 \citep{Choudhury2024}). All these properties can be directly mapped to the EoS of cold and $\beta$-equilibrated dense matter, even if they are not yet sufficient to firmly establish the composition of matter at high density.
The expected sensitivity for the third generation (3G) GW detectors (Cosmic Explorer \citep{Abbott_2017CE, reitze2019cosmicexploreruscontribution, CE2021} and Einstein Telescope \citep{Punturo2010, Hild2011,  Branchesi2023}) promises a large increase in the number of BNS merger detections and in the precision of the compact-object properties \citep{Abbott20173G, Hild2011}. The new Advanced Telescope for High-ENergy Astrophysics (ATHENA) mission \citep{ATHENA2013} is expected to detect soft X-rays emitted by millisecond pulsars, providing unprecedented spectral and time resolution to determine the atmosphere composition, mass, and radius of NSs, including for sources that are currently considered faint. An increase in the number of observed sources and in the accuracy of NS macroscopic parameter measurements will significantly improve constraints on dense-matter properties.

In addition to constraining the EoS, it may also be possible to constrain nuclear empirical parameters (NEPs; \citep{Myers1969}), such as the nuclear saturation density, the symmetry energy, or the incompressibility, with future NS observations. 
\citet{Iacovelli2023}  assessed the extent to which NEPs can be constrained using simulated GW data in the context of 3G detector sensitivity, and an EoS prior based on the meta-modeling approach (see \citet{Margueron2018} and reference therein). They highlight a degeneracy in the determination of NEPs that manifests itself as multi-peaked posterior distributions, even when informed by a large number of precise astrophysical detections. Other studies reach similar conclusions (e.g., \citet{Imam2024, Wouters2025, Zhu2025}).

In this paper, we discuss the performance of EoS and NEP inference using the semi-agnostic EoS construction presented in \citet{Davis2024}, that is, a meta-modeling approach coupled with a polytrope extension at high density. 
Section~\ref{sec:methods} details the methods used in this paper: we first introduce the semi-agnostic EoS model in Sect.~\ref{sec:methods/EoS}, present brief details on the NS macroscopic parameter modeling in Sect.~\ref{sec:methods/astro}, and describe the simple approach taken to simulate high-precision detections with a noncorrelated bivariate Gaussian distribution in Sect.~\ref{sec:methods/simu_data}. 
Section~\ref{sec:results} presents the results of our analysis. We describe the EoS prior sets used for the inference, their corresponding astrophysical parameters, and details of the NEPs in Sect.~\ref{sec:results/priors}, and discuss the NEP constraints obtained from simulated data in Sect.~\ref{sec:results/infer}. 
We provide additional discussion and conclusions in Sects.~\ref{sec:discussions} and \ref{sec:conclusion}, respectively.

\section{Methods} \label{sec:methods}

\subsection{Semi-agnostic EoS model} \label{sec:methods/EoS}

We constructed the EoS following the semi-agnostic (SA) approach discussed in \citet{Davis2024}: in the low-density regime, $n = [10^{-11}{\rm fm^{-3}} ; n_{\rm match}]$, the EoS is based on the meta-model introduced in \citet{Margueron2018},
and in the high-density regime, $n = [n_{\rm match} ; n_{\rm max}]$, the EoS is based on piecewise polytropes. We generated the different sets used in this paper with the Crust (Unified) Tool for Equation-of-state Reconstruction (\texttt{CUTER})\footnote{A publicly available version of  \texttt{CUTER}is accessible on the Zenodo repository at \url{https://doi.org/10.5281/zenodo.10781538} \citep{Davis2024, Davis2025}}. 
We refer to \citet{Davis2024} and references therein for a detailed description of the SA model, and briefly review its construction in the following sections.

\subsubsection{Low-density construction: The meta-model} \label{sec:methods/EoS/MM}

We calculated the zero-temperature and $\beta$-equilibrated low-density EoS of homogeneous matter composed of neutrons $n$, protons $p$, electrons $e$, and muons $\mu$ ($npe\mu$ matter) using the nucleonic meta-model approach.
This model is a non-relativistic, flexible parametrization of the baryonic part of the energy density, under the hypothesis that baryonic matter consists solely of neutrons and protons, with respective masses $m_q$ and densities $n_q$ ($q = n,p$). The baryonic energy density is given by
\begin{align}
    \epsilon_{\rm nuc}(n, \delta) &= (m_n n_n + m_p n_p)c^2  + \epsilon^*_{\rm FG}(n, \delta) \nonumber \\
    & \hspace{1.5cm}+ \epsilon_{\rm is}(n) + \epsilon_{\rm iv}(n) \delta^2 \;,
\end{align}
where $c$ is the speed of light, ${n = n_n + n_p}$ is the baryon density, and ${\delta = (n_n - n_p)/n}$ is the isospin asymmetry. We denote the interaction part of the baryonic energy density for symmetric matter (composed of an equal number of neutrons and protons) by $\epsilon_{\rm is}$ and refer to it as the isoscalar energy.
The symmetry energy density (or isovector energy) quantifies the energy required by the system to transform isospin-symmetric matter into pure neutron matter.
We denote the interaction contribution to the symmetry energy by $\epsilon_{\rm iv}$, while the kinetic term provides a first-order correction to the parabolic expansion in $\delta$ and follows the functional form of a free Fermi gas \citep{Margueron2018}:
\begin{equation}
    \epsilon^*_{\rm FG}(n, \delta) = \epsilon^*_{\rm FG}(n, \delta;  n_{\rm sat}, m^*_{\rm sat}, \Delta m^*_{\rm sat}) \;.
\end{equation}
The nuclear saturation density, denoted \nsat, is the baryon density at the nuclear saturation point, that is, the density defined by the minimum value of the energy per baryon in symmetric matter. We denote the Landau nucleon effective mass at nuclear saturation density in symmetric matter by $m^*_{\rm sat}$.
The effective mass isospin splitting (or isosplit), denoted $\Delta m^*_{\rm sat}$, is defined at nuclear saturation as the difference between the neutron and proton Landau effective mass in pure neutron matter (see Eq.~(6) of~\citet{Margueron2018}).

Both the isoscalar and isovector terms of the baryonic energy densities, $e_{\rm is}$ and $e_{\rm iv}$, are formulated as a fourth-order truncated Taylor expansion around \nsat.
Each order of the Taylor expansion is associated to a different NEP that also contains contributions from the density dependence of the effective masses.
We denote the ensemble of isoscalar and isovector NEPs by $X_{\rm is}$ and $X_{\rm iv}$, respectively, where
\begin{align}
    \{X_{\rm is}\} &= \{E_{\rm sat}, K_{\rm sat}, Q_{\rm sat}, Z_{\rm sat} \} \;, \\
    \{X_{\rm iv}\} &= \{E_{\rm sym}, L_{\rm sym}, K_{\rm sym}, Q_{\rm sym}, Z_{\rm sym} \} \;.
\end{align}
NEPs are defined at saturation density, with $E$ denoting the energy, $L$ the slope of the energy, $K$ the incompressibility, $Q$ the skewness, and $Z$ the kurtosis; the definition of the nuclear saturation point implies ${L_{\rm sat}=0}$. The isoscalar and isovector baryonic energy densities depend on $X_{\rm is}$ and $X_{\rm iv}$, respectively, as
\begin{align}
    \epsilon_{\rm is}(n) &= \epsilon_{\rm is}(n; n_{\rm sat}, \{X_{\rm is}\}, m^*_{\rm sat}, \Delta m^*_{\rm sat},
    b)\;, \\
    \epsilon_{\rm iv}(n) &= \epsilon_{\rm iv}(n; n_{\rm sat}, \{X_{\rm iv}\}, m^*_{\rm sat}, \Delta m^*_{\rm sat},
    b).
\end{align}
For details on the explicit formulas of $\epsilon_{\rm is}$ and $\epsilon_{\rm iv}$, we refer to Eqs.~(16)-(24) of~\citet{Davis2024} or to Eqs.~(22)-(31) of~\citet{Margueron2018}. 
The parameter $b$ enters the correction term for the potential energy that ensures the correct zero-density limit (see Eq.~(37) in \citet{Margueron2018}).
It has a negligible influence on the results presented in this paper.

Because the meta-model parametrizes only the baryonic contribution to the homogeneous-matter energy density, we added the contribution of leptons to build the EoS of $npe\mu$ matter.
For the inhomogeneous matter in the crust (at zero temperature), We used the compressible liquid drop model for the ions. In this approach, the cluster bulk energy (described within the meta-model) is complemented by the Coulomb interaction and the residual interface interaction between the nucleus and the surrounding dilute nuclear-matter medium, as described in \citet{Carreau2019, Davis2024}. 
We determined the parameters entering the cluster surface energy to reproduce the experimental nuclear masses in the 2020 Atomic Mass Evaluation (AME) table \citep{ame2020}.
We then obtained the unified EoS variationally, under the constraints of baryon number conservation and charge neutrality. 
We calculated the pressure $P$ from the energy density $\epsilon(n)$ using standard thermodynamic relations. 
We checked causality a posteriori for the meta-model EoS: we rejected models that present supra-luminal sound speeds before $n_{\rm match}$. 

At lower densities, we implemented the chiral effective field theory ($\chi$-EFT) constraint as detailed in \citet{Davis2024}: for each (meta-)model realization (that is, for each nuclear-parameter set sampled from the intervals in Table~\ref{tab:nuc_param}), we computed, for different densities $n$ in the interval ${n \in [\min(n_{\chi \rm data}), \min(n_{\rm match}, \max(n_{\chi \rm data}))]}$, the neutron-matter energy per particle $e_{\rm nuc}(n,\delta=1)$ and compared it with $\chi$-EFT calculations. We take these as a ``conflated'' band following \citet{Huth2022} (see their Fig.~4), in contrast to \citet{Davis2024}, who used the calculations of \citet{Drischler2016}. The probability density is Heaviside-like; that is, it is flat within the conflated band, enlarged by $5\%$ on each side, and zero otherwise. \citet{Scurto2024} show that this procedure yields posteriors in agreement with those obtained by considering the conflated band as a $90\%$ confidence interval, adopting a flat probability inside the band and Gaussian tails outside to account for the remaining $\pm5\%$ (as also done in \citet{Montefusco2025}).

We abandoned the meta-model beyond $n_{\rm match}$ in favor of a more agnostic approach because: 
(i) the Taylor expansion around nuclear saturation density $n_{\rm sat}$ fails for large density values \citep{Margueron2018}, and
(ii) the purely nucleonic degrees of freedom on which the meta-model is based might not be correct in the high-density regime, as the core of NSs may contain more exotic particles, such as hyperons (see e.g., \citet{Vidana2016, Oertel2016}) or deconfined quarks (see e.g., \citet{Baym2018}).

\subsubsection{High-density construction: Piecewise polytropes} \label{sec:methods/EoS/PP}

Beyond $n_{\rm match}$, we used the agnostic formulation of piecewise polytropes (see~\citet{Read2009}); that is, the pressure and rest-mass density $\rho = n m_n$ (with $m_n$ the nucleon mass) are related according to the formula $P=\kappa \rho^{\gamma}$, where $\kappa$ is the polytropic constant and $\gamma$ the adiabatic index. 
We used $N=5$ polytropes defined by ${\{\gamma_i, \kappa_i, \rho_{i-1}^{\rm tr} \}}$, where ${\rho_{i-1}^{\rm tr}}$ is the transition density between the polytropes and ${i \in [1;N]}$. 

We matched the first polytrope to the low-density meta-model EoS at ${\rho^{\rm tr}_0= \rho_{\rm match} = m_n n_{\rm match}}$ and ensured pressure continuity at all transition densities between the polytropes. 
Therefore, the high-density EoS is entirely known if we specify $n_{\rm match}$, five adiabatic indices, and four transition densities $\rho_{i-1}^{\rm tr}$. We computed $P(n)$ and $P(\epsilon)$ in the density range ${[n_{\rm match}; {\rm max}(n_{\rm cs}, 20n_0)]}$, with ${n_0=0.16}$\fmc and $n_{\rm cs}$ the maximum baryon density that ensures causality of the polytropic extension, because polytropes are not causal by construction.
Therefore, the number of EoS polytropes may be less than five if causality is broken before the subsequent transition density is reached.

The matching procedure between the meta-model and the polytropic EoS and between two polytropes follows the methodology described in \citet{Davis2024}. We chose the constant $\kappa$ of each polytrope to ensure continuity between the different EoS pieces (including at $n_{\rm match}$). We calculated the energy density of the polytrope using the first law of thermodynamics (at zero temperature), and  we chose the integration constant (see e.g., \citet{Read2009}) at the matching points such that continuity and thermodynamic consistency are satisfied.

\citet{Komoltsev2022} propose a method to link the NS EoS to the perturbative quantum chromodynamic (pQCD) regime. This allows the exact pQCD results to be integrated toward lower densities, providing an additional constraint 
on the EoS in the high-density regime. However, \citep{Komoltsev2024} show that
the termination density, up to which the EoS modeling is performed in an inference setup, strongly affects the constraining power of the pQCD input.
In particular, if this density is chosen as the central density $n_{\rm TOV}$ at the maximum mass $M_{\rm TOV}$ of each EoS model, the impact of the pQCD constraint is marginal.
Because we are only interested in determining the EoS along the stable NS branch, and not in studying the QCD physics above $n_{\rm TOV}$, we consider that our EoS modeling based on piecewise polytropes at high density cannot be safely extrapolated at densities above $n_{\rm TOV}$. Within this more conservative viewpoint, we did not implement the pQCD constraint.

\subsubsection{Equation-of-state prior sets} \label{sec:methods/EoS/priors}

\begin{table}[]
    \centering
    \caption{Sampling ranges for the meta-model nuclear parameters $X$.}
    \begin{tabular}{|c||c|c|}
        \hline
        Parameter $X$ & $X_{\rm min}$ & $X_{\rm max}$ \\ \hline \hline
        $n_{\rm sat}$ [\fmc] & 0.14 & 0.17 \\ \hline
        $E_{\rm sat}$ [MeV] & -17.0 & -15.0 \\ \hline
        $K_{\rm sat}$ [MeV] & 190 & 290 \\ \hline
        $Q_{\rm sat}$ [MeV] & -1000 & 1000 \\ \hline
        $Z_{\rm sat}$ [MeV] & -3000 & 3000 \\ \hline
        $E_{\rm sym}$ [MeV] & 26 & 38 \\ \hline
        $L_{\rm sym}$ [MeV] & 10 & 80 \\ \hline
        $K_{\rm sym}$ [MeV] & -400 & 200 \\ \hline
        $Q_{\rm sym}$ [MeV] & -2000 & 2000 \\ \hline
        $Z_{\rm sym}$ [MeV] & -5000 & 5000 \\ \hline
        $m^*_{\rm sat}/m$ & 0.6 & 0.8 \\ \hline
        $\Delta m^*_{\rm sat}/m$ & 0.0 & 0.2 \\ \hline
        $b$ & 1 & 10 \\ \hline
    \end{tabular}
    \tablefoot{The mean effective mass and effective isosplit are normalized to $m=\frac{m_n+m_p}{2}$.}    
    \label{tab:nuc_param}
\end{table}

We sampled NEPs along a uniform distribution over the intervals presented in Table~\ref{tab:nuc_param} to compute the low-density EoS (${ n \in [10^{-11}\text{\fmc};n_{\rm match}]}$). 
The prior range for the NEPs is similar to that used in \citet{Dinh2021Uni, Davis2024}, but we slightly enlarged it for some NEPs to accommodate the injection values used in our study. 
We chose these ranges to be wide enough to be fully consistent with current nuclear phenomenology, as reviewed in \citet{Margueron2018} (see also \citet{Oertel2017, LeFevre2016, Danielewicz2002}).
These prior intervals are also close to those used in \citet{Patra2023}, who performed a Bayesian analysis that focused on the impact of symmetry-energy parameters. In contrast to that work, we used slightly larger ranges for the $Q_{\rm sat,sym}$ parameters to allow more flexibility, as these parameters are experimentally unconstrained.
\citet{Tsang2020} also used a meta-model-based EoS, but kept $E_{\rm sat}$ and $n_{\rm sat}$ fixed, whereas we varied these parameters. Although these low-density parameters are not very influential for global astrophysical observables, we consider their uncertainty non-negligible, as nuclear physics experiments cannot yet determine their value with high precision.

Moreover, we used a uniform initial distribution for the parameters (see Table~\ref{tab:nuc_param}) instead of a Gaussian distribution to adopt a more agnostic approach and avoid biasing the analysis.
However, fitting the surface parameters to the experimental nuclear masses in the compressible liquid-drop model, as well as the requirement that the EoS models are causal, thermodynamically consistent, and able to produce a consistent crust, reduces the NEP space of the actual ``prior.'' 
As a result, the prior NEP distribution used in the Bayesian inference with astrophysical data is not necessarily flat (see e.g., \citet{Dinh2021Uni}). 

We sampled the adiabatic indices $\gamma$ and transition densities $\rho^{\rm tr}$ for the piecewise polytropes along a uniform distribution over the intervals $\gamma \in [0,8]$ and $[\rho_{\rm match},10 n_0 m_n]$, respectively.
We prepared four different sets using the \texttt{CUTER} code:
\begin{itemize}
  \item \expa: follows the semi-agnostic approach discussed in this section, with the meta-model construction extending up to ${n_{\rm match} = n_0 = 0.16}$\fmc and piecewise polytropes beyond ${n_{\rm match}}$. This set is informed solely by nuclear experiments, which guided the limits of the NEPs priors.
  \item \expb: similar to \expa\; but with piecewise polytropes matched to the meta-model at ${n_{\rm match} = 2n_0 = 0.32}$\fmc  .
  \item \chiset: similar to \expa\; ($n_{\rm match} = n_0$), with the addition of $\chi$-EFT constraints imposed for ${n \in [\min(n_{\chi \rm data});n_0]}$.
  \item \mmset: the low-density meta-model is extended at high density instead of using piecewise polytropes. We imposed $\chi$-EFT constraints  for ${n \in [\min(n_{\chi \rm data})-\max(n_{\chi \rm data})] = [0.02-0.2]}$\fmc.
\end{itemize}

Each set comprised $5\times10^4$ EoSs and was used as a prior for Bayesian inference with astrophysical data.

\subsection{Neutron-star astrophysical parameters} \label{sec:methods/astro}

For each EoS in the sets, we modeled the NS total mass $M$, total radius $R$, and dimensionless tidal deformability $\Lambda$ in the framework of general relativity. We computed the mass-radius sequence for each EoS  by solving Einstein's equations for a static metric (Tolman-Oppenheimer-Volkoff equations \citep{Tolman1939, Oppenheimer1939}), while varying the central energy density of the star.

The dimensionless tidal deformability is defined as
\begin{equation}
  \Lambda = \frac{2}{3} k_2(r=R) \bigg( \frac{Rc^2}{GM}\bigg)^5 \;,
\end{equation}
where $k_2(r)$ is the second-order tidal Love number. 
The quantity $k_2$ is the solution of ordinary differential equations presented in \citet{Hinderer2008}, which we solved simultaneously with the Tolmann-Oppenheimer-Volkov equations.
This generated, for each EoS, a set of NS modeled sources with their associated $M, R,$ and $\Lambda$.

\subsection{Simulated astrophysical data} \label{sec:methods/simu_data}

\begin{table*}
    \centering
    \caption{Nuclear and astrophysical properties of the injection EoSs.}
    \begin{tabular}{|c||c|c|c|}
        \hline
                            & RG(SLy2)              & PCP(BSk24)               & GPPVA(NL3-$\omega \rho$)                \\ \hline \hline
                            & \multicolumn{3}{c|}{Astro. for $M = (1.0, 1.4, 2.0)$\mdot}                \\ \hline
    $n_c$ [\fmc]            & (0.407, 0.530, 0.941) & (0.331, 0.407, 0.589)    & (0.256, 0.293, 0.358)     \\ \hline
    $P_c$ [MeV/fm$^3$]      & (37.8, 82.7, 428.7)   & (28.6, 55.5, 165.8)      & (20.4, 35.0, 73.6)        \\ \hline
    $R$ [km]                &  (11.91, 11.76, 10.71)& (12.46, 12.58, 12.31)    & (13.52, 13.81, 14.09)     \\ \hline
    $\Lambda$               & (2342.4, 310.0, 11.3) & (3279.2, 518.9, 40.6)    & (5081.7, 944.3, 119.8)    \\ \hline \hline
                            & \multicolumn{3}{c|}{Nuclear empirical parameters}                         \\ \hline
    n$_{\rm sat}$ [\fmc]    & 0.161                 & 0.1578                   & 0.148                                \\ \hline
    E$_{\rm sat}$ [MeV]     & -15.99                 & -16.048                  & -16.24                               \\ \hline
    K$_{\rm sat}$ [MeV]     & 229.92                & 245.5                    & 270.0                                \\ \hline
    Q$_{\rm sat}$ [MeV]     & -                     & -274.5                    & -198.0                                \\ \hline
    Z$_{\rm sat}$ [MeV]     & -                     & 1184.15                  & -                                    \\ \hline
    E$_{\rm sym}$ [MeV]     & 32.0                  & 30.0                     & 31.7                                 \\ \hline
    L$_{\rm sym}$ [MeV]     & 47.46                 & 46.4                     & 55.0                                 \\ \hline
    K$_{\rm sym}$ [MeV]     & -115.13               & -37.6                    & -8.0                                 \\ \hline
    Q$_{\rm sym}$ [MeV]     & -                     & 710.86                   & -                                    \\ \hline
    Z$_{\rm sym}$ [MeV]     & -                     & -4031.3                  & -                                    \\ \hline \hline
    \end{tabular}
    \tablefoot{Central baryon density $n_c$, central pressure $P_c$, radius $R$,
    and tidal deformability $\Lambda$ for $M=1.0,1.4,2.0$\mdot, together with the available nuclear empirical parameters of the underlying nuclear models for the RG(SLY2), PCP(BSK24), and GPPVA(NL3$\omega\rho$) EoSs.}
    \label{tab:injection_data}
\end{table*}

We simulated data on $M$, $R$ and $\Lambda$ and studied the recovery of the EoS and NEPs through inference. We used three unified\footnote{The core and the crust of a unified EoS are calculated consistently using the same nuclear model.} injection nucleonic EoSs  available on the CompOSE database \citep{Typel2022}  to simulate the data: RG(SLY2) \citep{Gulminelli2015}, PCP(BSK24) \citep{Pearson2018}, and GPPVA(NL3$\omega\rho$) \citep{Grill2014}. Table~\ref{tab:injection_data} lists the nuclear and astrophysical properties of the injection EoSs.

Unlike \citet{Boudon2025} and \citet{Iacovelli2023}, we did not perform a full population study to simulate the astrophysical data in our study. 
Rather, we focused on simulating NS sources detected with high precision and at the extremes of the plausible mass range for astrophysical NSs to test the limits of EoS and NEP inference capabilities.
For each injection EoS, we assumed the detections of the mass, radius, and tidal deformability for: 
\begin{itemize}
    \item ten NS sources with uniformly distributed masses $M_{\rm source} \in [0.9-1.1]$\mdot. We refer to this ensemble of NS sources as ``low-mass'' sources.
    \item ten NS sources with uniformly distributed masses $M_{\rm source} \in [1.8-2.0]$\mdot. We refer to this ensemble of NS sources as ``high-mass'' sources.
\end{itemize}
For each NS source, we simulated the measurement of astrophysical parameters using an uncorrelated bivariate Gaussian distribution centered on $\big(M_{\rm source}, X^{\rm inj}(M_{\rm source}) \big)$ with $X=\{R, \Lambda\}$ and $X^{\rm inj}$ determined by the injection EoS. We generate the ensemble of $K$ simulated data points $\{ M^K_{\rm source}, X^K(M_{\rm source}) \}$ with the standard deviation $(\sigma_M, \sigma_X) = (M_{\rm source},X^{\rm inj}(M_{\rm source}) \times \delta E $, with $\delta E=1\%$. 

We acknowledge that the small assumed error $\delta E$ does not accurately reflect the precision offered by the next generation of detectors, nor the fact that the different observables are affected by very different systematics; moreover, the bivariate Gaussian distribution does not properly account for correlations between NS macroscopic parameters.
We used this simplistic approach to test the capabilities of the semi-agnostic EoS sets to recover the injection values.

\subsection{Astrophysical inference}

To determine the impact of the simulated observations on the EoS, we weighted each EoS in our sample according to its likelihood under the bivariate Gaussian observation. The astrophysical implications of each EoS are captured by the family of possible stars with properties $\{M, X_{\rm eos}(M)\}$.

Using the probability density function method embedded in the package \textsc{stats.multivariate\_normal} of the \textsc{scipy} library \cite{Scipy2020}, we computed the likelihood $\mathcal{L}(d_{\rm source}|\mathrm{eos})$ of measuring each source for each EoS in the set. We generated the set of $N$ possible stars associated with the EoS, $\{ M^K_{\rm eos}, X^K(M_{\rm eos}) \}$, with uniformly spaced sampling in mass, effectively assuming a flat population model. The likelihood is then given by
\begin{equation}
    \mathcal{L}(d_{\rm source}|\mathrm{eos}) = \frac{1}{N} \sum_{K=1}^N p(d_{\rm source}|M^K_{eos},X(M^K_{\rm eos})),
\end{equation}
where $p(d_{\rm source}|M,X)$ denotes the multivariate normal likelihood distribution associated with each simulated observation. This method is similar to that used in \cite{Essick:2019ldf}.
Finally, we determined the likelihood $\mathcal{L}(d_{\rm population}|\mathrm{eos})$ for each EoS in the set given the ten NS detections by multiplying the likelihood per source.

By weighting each sample in our initial prior EoS distribution by the likelihood of the EoS under the set of simulated observations, we can determine the posterior distribution of all quantities captured by the EoS distribution (such as the distribution of the associated NEPs) following Bayes theorem.

\section{Results} \label{sec:results}

\subsection{Priors} \label{sec:results/priors}

\subsubsection{Equation of state}

\begin{figure}
    \centering
    \resizebox{\hsize}{!}{\includegraphics{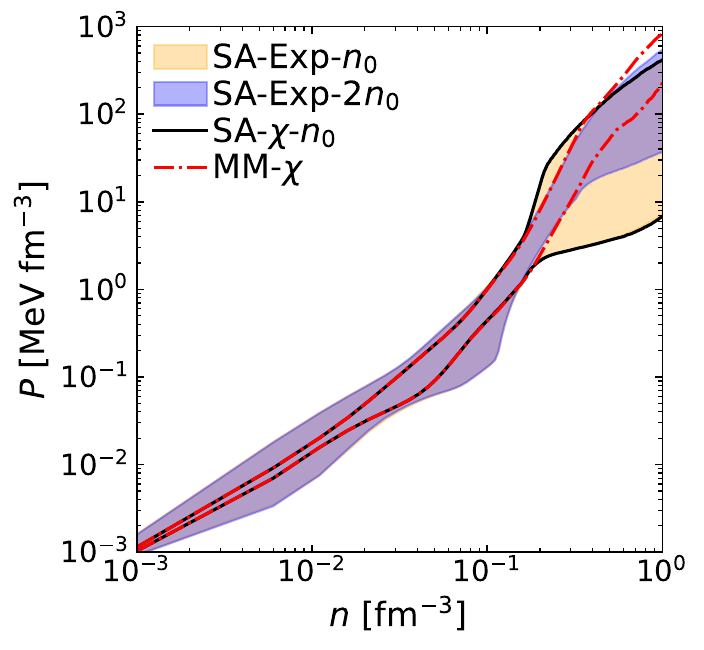}}
    \caption{Pressure $P$ as a function of the baryon density $n$ (90 \% contours) for the EoS sets discussed in this paper.}
    \label{fig:eos_set_comparison/compare}
\end{figure}

\begin{table*}
    \centering
    \caption{90\% contour pressure range $\delta P(n)$ in MeV/fm$^3$. }
    \begin{tabular}{|c|c|c|c|c|c|c|}
        \hline
         Set & $\delta P(n_0/10)$ & $\delta P(n_0/2)$ & $\delta P(n_0)$ & $\delta P(2 n_0)$ & $\delta P(4n_0)$ & $\delta P(6n_0)$ \\ \hline \hline
        \expa         & 0.05 & 0.59 & 2.8 & 64.5 & 223.2 & 384.9  \\ \hline
        \expb         & 0.05 & 0.58 & 2.7 & 39.6 & 240.8 & 480.7  \\ \hline
        \chiset       & 0.01 & 0.30 & 2.6 & 64.8 & 226.0 & 391.5  \\ \hline
        \mmset        & 0.01 & 0.30 & 2.4 & 41.5 & 239.0 & 575.4 \\ \hline
    \end{tabular}
    \tablefoot{The quantity $\delta P(n)$ is calculated at $n_0/10$, $n_0/2$, $n_0$, $2 n_0$, $4n_0$, and $6n_0$ for the EoS sets discussed in this paper.}
    \label{tab:deltaP_sets}
\end{table*}

We present the 90\% contours for the pressure-density distribution of the four EoS sets generated with \texttt{CUTER} (see Section~\ref{sec:methods/EoS}) in Fig.~\ref{fig:eos_set_comparison/compare}, and we present the corresponding pressure range $\delta P$ (on the 90\% contours) explored by each set in Table~\ref{tab:deltaP_sets} for various values of the baryon density.

For $n\in [10^{-3};0.1]$\fmc, the $\chi$-EFT filters significantly constrain the pressure range explored by the meta-model. For $n\in[0.1-0.16]$\fmc, all four sets approximately converge in terms of $\delta P$, with or without $\chi$-EFT filters applied. Although the $\chi$-EFT calculations considered extend up to ${n=0.2}$\fmc, the error bar on the $\chi$-EFT data for $n > 0.1$\fmc is significant, and the pressure range is not constrained beyond the  nuclear experimental data intervals from which the meta-model EoSs are sampled.

For ${n \in [0.16 ; 0.32]}$\fmc, the sets \mmset (full meta-model with $\chi$-EFT constraints) and \expb (semi-agnostic with ${n_{\rm match} = 0.32}$\fmc without $\chi$-EFT constraints) disfavor high and low pressures compared to \expa: the meta-model construction with its nucleonic hypothesis presents a narrow $\delta P$ in this density regime compared to piecewise polytropes, which can explore a pressure range up to two times larger. For the semi-agnostic sets, extending the meta-model up to ${n_{\rm match}=0.32}$\fmc (\expb) results in tighter pressure contours in ${n \in [n_0; 2n_0]}$.

For $n > 2n_0$, the sets \expa, \expb, and \chiset yield a larger $\delta P$ than \mmset. In this density regime, the semi-agnostic sets allow for significantly softer core EoSs, while \mmset presents particularly stiff EoS contours. The semi-agnostic sets can mimic the presence of non-nucleonic degrees of freedom in the core of the star, producing an overall softening of the EoS.

\subsubsection{Neutron-star astrophysical parameters}
\label{sec:results/NSmacro}

\begin{figure}
    \centering
    \resizebox{\hsize}{!}{\includegraphics{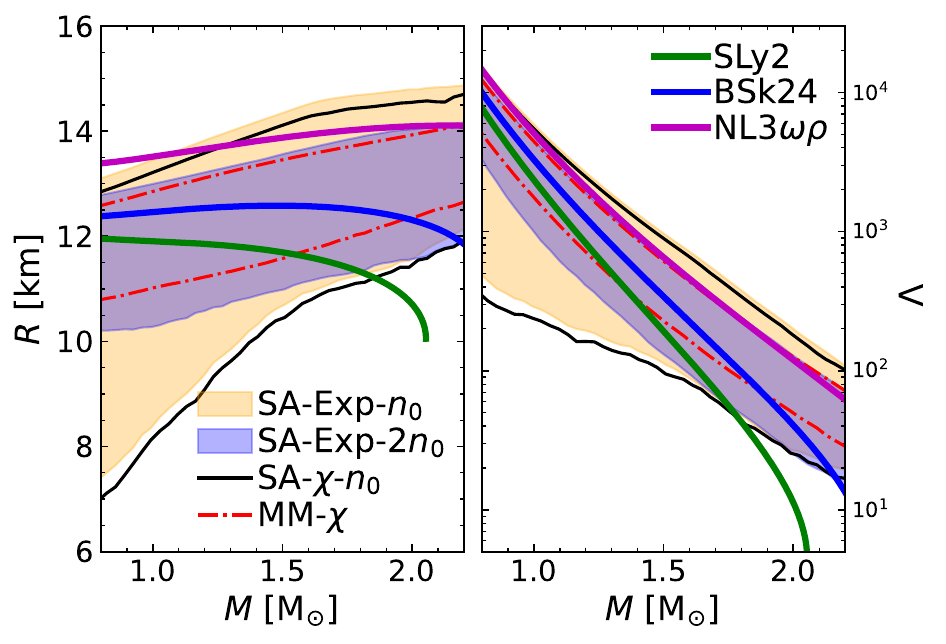}}
    \caption{Radius $R$ and tidal deformability $\Lambda$ as a function of the mass $M$ (90 \% contours) for the EoS sets discussed in this paper.  The $M(R)$ and $M(\Lambda)$ sequence for EoSs RG(SLY2), PCP(BSK24), and GPPVA(NL3$\omega\rho$) (used as injection EoSs) are also shown in green, blue, and magenta, respectively.}
    \label{fig:obs_set_comparison/compare_mrl}
\end{figure}

\begin{table*}
    \centering
    \caption{90\% contour radius and tidal deformability ranges.}
    \begin{tabular}{|c|c|c|c||c|c|c|}
        \hline
        Set & $\delta R$(\mdot) & $\delta R$(1.4\mdot) & $\delta R$(2\mdot) & $\delta \Lambda$(\mdot) & $\delta \Lambda$(1.4\mdot) & $\delta \Lambda$(2\mdot) \\ \hline \hline
        \expa         & 5.06 & 3.87 & 3.23 & 5630 & 1194 & 174    \\ \hline
        \expb         & 2.72 & 2.57 & 2.47 & 3568 & 694  & 103    \\ \hline
        \chiset       & 5.05 & 3.86 & 3.10 & 5244 & 1107 & 154    \\ \hline
        \mmset        & 1.84 & 1.90 & 1.57 & 2685 & 548  & 76     \\ \hline
    \end{tabular}
    \tablefoot{We calculated $\delta R(M)$ (in km) and  $\delta \Lambda(M)$ at $M = 1, 1.4$, and $2$\mdot for the EoS sets discussed in this paper. }
    \label{tab:deltaRML_sets}
\end{table*}

\begin{figure}
    \centering
    \resizebox{\hsize}{!}{\includegraphics{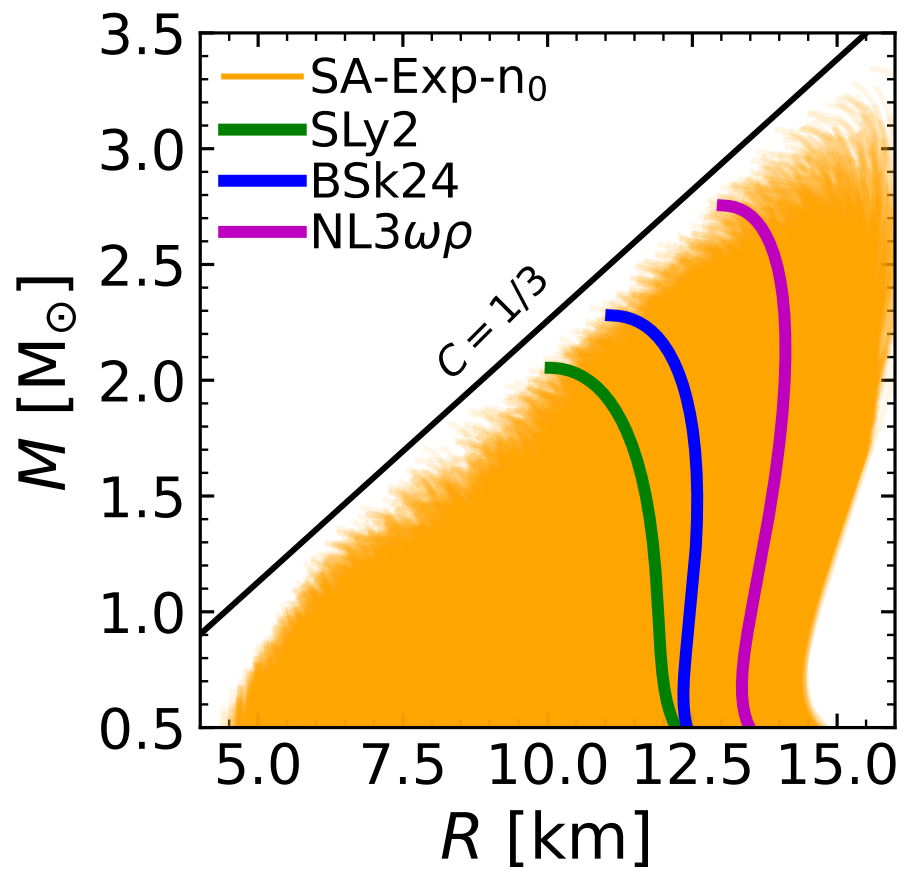}}
    \caption{Mass-radius relation for all EoSs in the \expa set and for the injection EoSs RG(SLY2), PCP(BSK24) and GPPVA(NL3$\omega\rho$) shown in green, blue, and magenta, respectively. The compactness limit $C=1/3$ is shown in black (see text for details).}
    \label{fig:full_MR-SA-Exp_n0}
\end{figure}

Figure~\ref{fig:obs_set_comparison/compare_mrl} shows the 90\% mass-radius and mass-tidal deformability contours corresponding to the four EoS prior sets; we list the radius and tidal deformability ranges $\delta R$  and $\delta \Lambda$ (on the 90\% contours) explored by each set in Table~\ref{tab:deltaRML_sets} for various values of the total NS mass. We also show the $M(R)$ and $M(\Lambda)$ sequences for the three injection EoSs RG(SLY2), PCP(BSK24), and GPPVA(NL3$\omega\rho$) in Fig.~\ref{fig:obs_set_comparison/compare_mrl}. Additionally, we present the mass-radius relation for all individual EoSs in the \expa set in Fig.~\ref{fig:full_MR-SA-Exp_n0}, as well as the mass-radius sequence for the three aforementioned injection EoSs. This figure also shows the compactness limit, which is respected by all four EoS sets presented in this paper (see Fig.~\ref{fig:full_MR-all} in Appendix~\ref{sec:app/mr-ml}) and is discussed in Ref.~\cite{Rezzolla2026}. Of the four EoS sets discussed in this paper, \expa is the least informed by nuclear physics and explores the mass-radius plane widely, albeit nonuniformly (see the difference with the 90\% contours presented in Fig.~\ref{fig:obs_set_comparison/compare_mrl}). It covers all three injection EoSs up to their Tolmann-Oppenheimer-Volkoff (TOV) mass limits.

The full meta-model set, \mmset, leads to the tightest contours in both $M(R)$ and $M(\Lambda)$, as expected by the EoS contours presented in Fig.~\ref{fig:eos_set_comparison/compare}. The semi-agnostic \expb set presents the second tightest contours of the four sets, while the ranges explored by \expa and \chiset are quite similar. However, the \chiset set, informed by $\chi$-EFT data, prefers slightly smaller radii and tidal deformabilities (more compact objects).

The injection EoS GPPVA(NL3$\omega\rho$) falls slightly outside or on the border of the 90\% contours for all EoS sets in the range of very low mass NSs; this EoS is particularly stiff and leads to high radii and high tidal deformabilities. Similarly, the injection EoS RG(SLY2) steps outside of the 90\% contours of the astrophysical parameter priors (for all sets) when considering high mass NSs. Although the injection nuclear parameters (see Table~\ref{tab:injection_data}) are within the prior interval presented in Table~\ref{tab:nuc_param}, this does not guarantee a large statistical realization of similar EoSs and the corresponding astrophysical parameters in the semi-agnostic sets.

\subsubsection{Correlation between the pressure and nuclear parameters}\label{sec:results/prior/correlation}

\begin{figure*}
    \sidecaption
    \includegraphics[width=12cm]{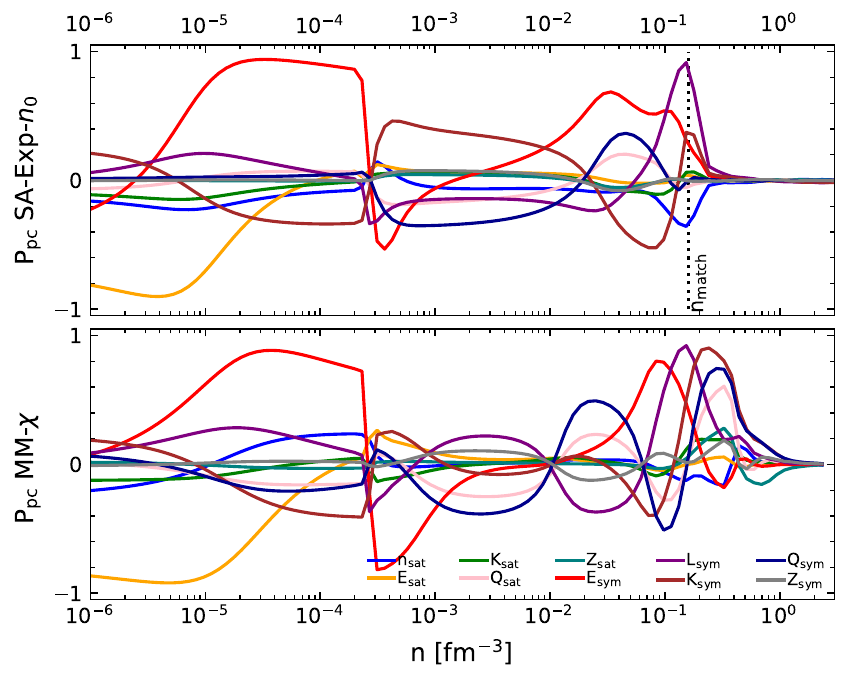}
    \caption{Pearson correlation factors between the $\beta$-equilibrated pressure, $P_{\rm pc}$, and the nuclear empirical parameters as a function of the baryon density, $n$, for the sets \expa and \mmset. For \expa, the matching density to polytropes is represented as a vertical dashed line.}
    \label{fig:pearson_density}
\end{figure*}

We present the Pearson correlation factors between NEPs and the $\beta$-equilibrated pressure ($P_{\rm pc}$) as a function of the baryon density for the sets \expa and \mmset in Fig.~\ref{fig:pearson_density}.

In the low-density part of the outer crust (${n\lesssim  10^{-5}}$\fmc), the nuclei are almost isospin symmetric and \esat dominates the contribution. The main pressure contribution in the outer crust comes from electrons: small values of \esat imply a small nuclear binding energy and small nuclei, which is why this parameter anticorrelates with the pressure.
In the high-density part of the outer crust (${10^{-5} \lesssim n\lesssim 4 \times 10^{-4} }$\fmc), the nuclei in the lattice become more neutron rich: the symmetric-matter energy correlates less with the pressure and \esym starts to dominate the contribution (and, to a lesser extent, also \ksym).

In the low-density part of the inner crust (${4 \times 10^{-4} \lesssim n \lesssim 10^{-2} }$\fmc), neutrons start dripping out of nuclei and contributing to the pressure, which is why the \esym correlation with pressure drops significantly right after the outer--inner crust transition.
The very low density of the neutron gas (much lower than the saturation density) also explains why the higher-order parameters in the expansion, \ksym and \qsym, take over the pressure contribution; as we progress deeper into the inner crust, the neutron-gas density increases, and \esym once again contributes significantly to the pressure. The $\chi$-EFT calculations have a strong impact on \ksym and \qsym in the low-density inner crust, acting as a decorrelating agent for those parameters.
In the density regime ${10^{-2} \lesssim n \lesssim 10^{-1} }$\fmc, the neutron gas reaches densities similar to those of the clusters, and \esym dominates the pressure contribution because the unbound neutrons are in a high-density environment and the nuclei are also very neutron-rich. This region contains the core--crust transition for most of the models, which explains the rapid evolution of many parameters.

In the lower-density part of the core, a strong correlation between the pressure and \lsym appears:
the pressure of pure neutron matter at nuclear saturation is solely determined by \lsym in the parabolic approximation to the symmetry energy (see \citet{Lattimer2014}).
The isovector parameters \lsym, \ksym, and \qsym successively become the most important contribution to the pressure as the density in the core increases, as per the Taylor expansion around saturation density on which the meta-model is based.
The correlation between the pressure and NEPs decreases rapidly beyond ${n=n_{\rm match}}$ for \expa: NEPs have minimal influence on the pressure supported by the polytropes, although the correlation extends beyond $n_{\rm match}$. This occurs because the polytropic constant of the first polytrope in the agnostic construction is connected to the meta-model pressure at the match (see Sect.~\ref{sec:methods/EoS/PP}).

For \mmset, even though the EoS is parametrized by the NEPs up to the highest densities, we observe a decrease and extinction of the correlation between the pressure and the nuclear parameters for ${n\gtrsim 0.4}$\fmc: different NEP combinations lead to the same pressure at very high density. This attests to an important degeneracy among the different NEPs that contribute to the $\beta$-equilibrated pressure, particularly between the isoscalar and isovector sectors. This is consistent with the findings of \citet{Iacovelli2023} and \citet{Mondal2022}. Several parameters are of comparable importance in determining the $\beta$-equilibrated pressure, and this degeneracy challenges the inference of NEPs based on the $\beta$-equilibrated pressure at high density.

\subsection{Inference with simulated data}
\label{sec:results/infer}

\subsubsection{Equation of state posteriors} \label{sec:results/inference/eos}

\begin{figure*}[!ht]
\centering
\begin{subfigure}{0.32\textwidth}
    \resizebox{\hsize}{!}{\includegraphics{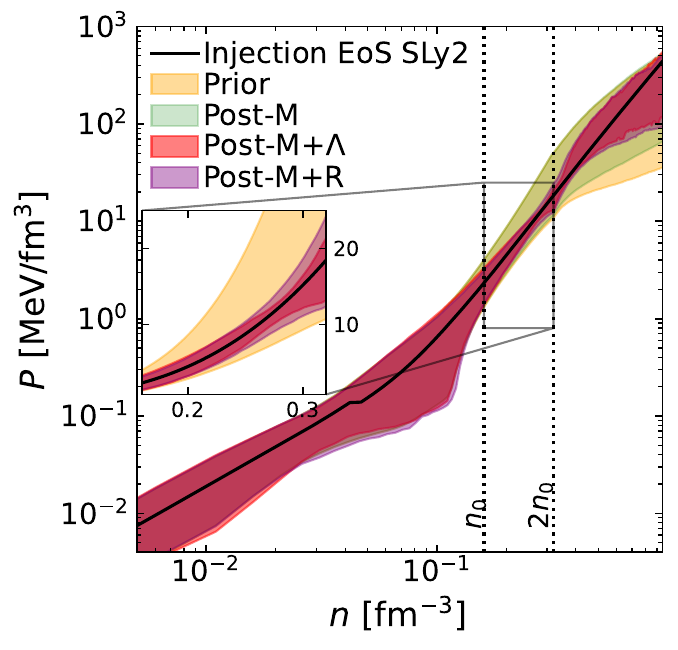}}
    \caption{RG(SLY2), low masses.}
    \label{fig:post_1pct_small_sly2}
\end{subfigure}
\hfill
\begin{subfigure}{0.32\textwidth}
    \resizebox{\hsize}{!}{\includegraphics{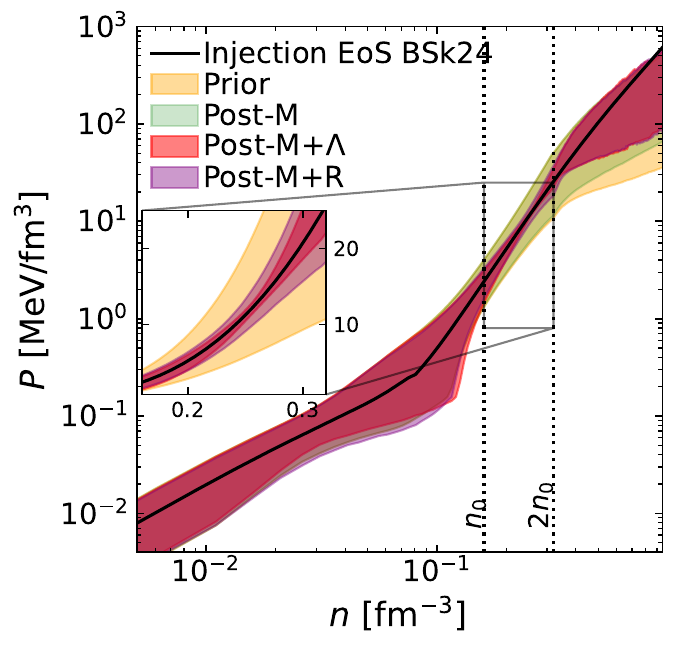}}
    \caption{PCP(BSK24), low masses.}
    \label{fig:post_1pct_small_bsk24}
\end{subfigure}
\hfill
\begin{subfigure}{0.32\textwidth}
    \resizebox{\hsize}{!}{\includegraphics{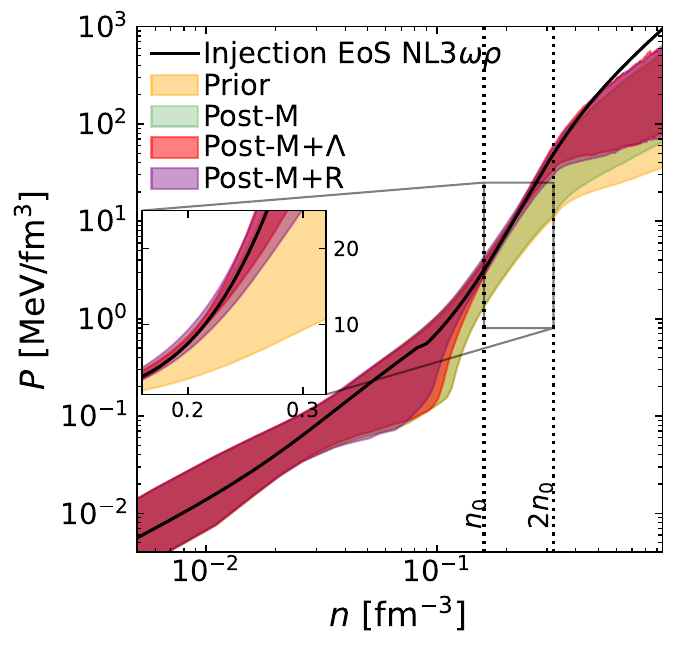}}
    \caption{GPPVA(NL3$\omega\rho$), low masses.}
    \label{fig:post_1pct_small_nl3}
\end{subfigure}
\hfill
\begin{subfigure}{0.32\textwidth}
    \resizebox{\hsize}{!}{\includegraphics{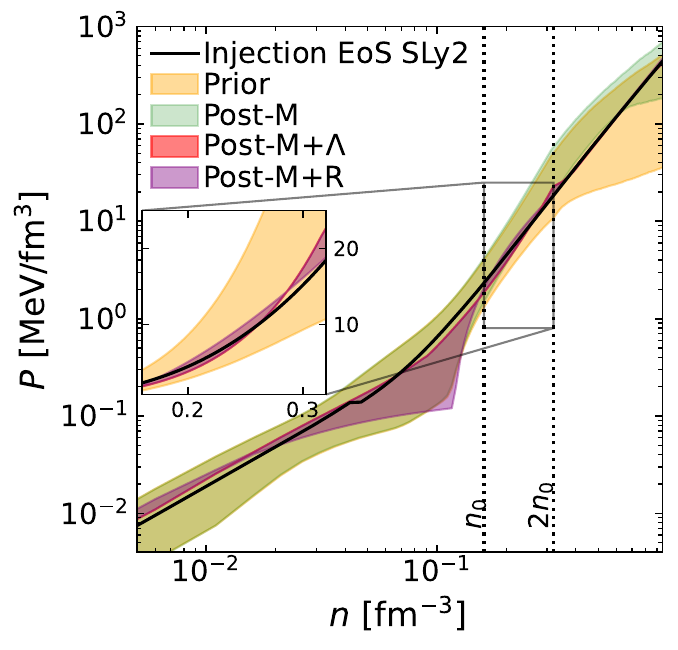}}
    \caption{RG(SLY2), high masses.}
    \label{fig:post_1pct_high_sly2}
\end{subfigure}
\hfill
\begin{subfigure}{0.32\textwidth}
    \resizebox{\hsize}{!}{\includegraphics{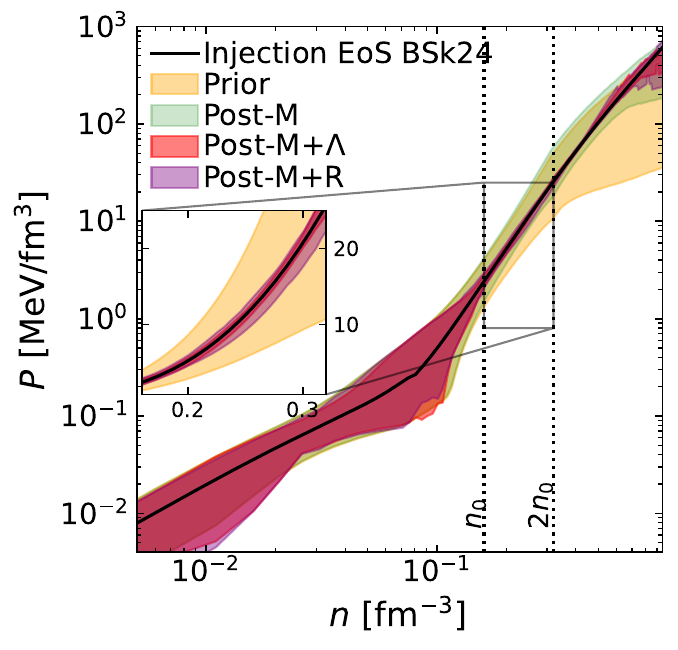}}
    \caption{PCP(BSK24), high masses.}
    \label{fig:post_1pct_high_bsk24}
\end{subfigure}
\hfill
\begin{subfigure}{0.32\textwidth}
    \resizebox{\hsize}{!}{\includegraphics{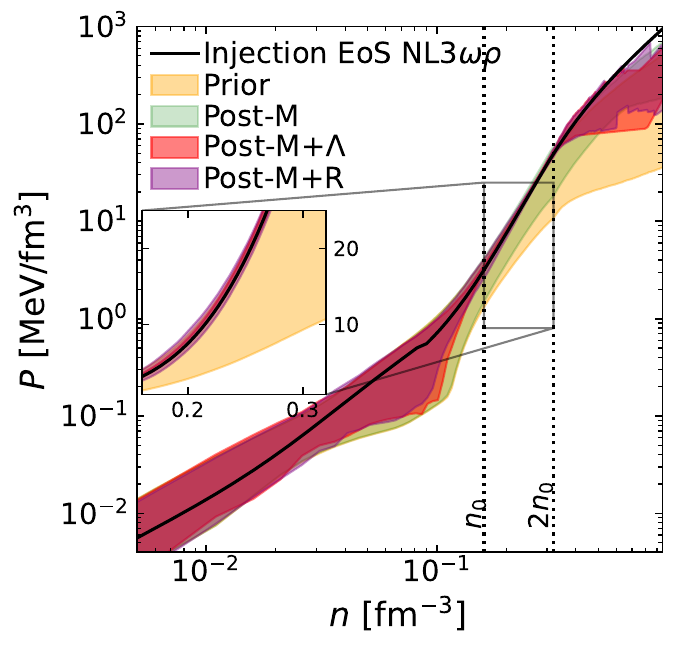}}
    \caption{GPPVA(NL3$\omega\rho$), high masses.}
    \label{fig:post_1pct_high_nl3}
\end{subfigure}
\caption{Pressure $P$ as a function of the baryon density $n$ (90 \% contours) for the prior \expb (orange) and the posterior informed by simulated detections of ten NSs with 1\% relative error in $M$ (green), $(M, \Lambda)$ (red), and $(M,R)$ (purple). The injection EoSs (RG(SLY2), PCP(BSK24), and GPPVA(NL3$\omega\rho$)) are shown in black.}
\label{fig:post_1pct}
\end{figure*}

Figure~\ref{fig:post_1pct} displays the EoS posteriors based on the \expb set and informed by simulated detections of $M$, $M-R$, and $M-\Lambda$ with 1\% precision for the three injection EoSs RG(SLY2), PCP(BSK24), and GPPVA(NL3$\omega\rho$), considering low-mass NS sources (top panels) and high-mass NS sources (bottom panels). We present the corresponding effective sample sizes of the posteriors informed by $(M, \Lambda)$ detections (results are similar for $(M,R)$ detections) in Table~\ref{tab:effective_sample_size} of Appendix~\ref{sec:app/size}.

The differences between the posteriors informed by $M$, $M-R$, or $M-\Lambda$ detections show that the detection of the mass alone (in green) only pushes the posterior towards high pressure values in the high-density regime, as expected. Indeed, stiffer EoSs are needed to counterbalance gravity for massive stars. Simulated $M-R$ (in purple) and $M-\Lambda$ (in red) detections provide tight constraints in the density regime $n \in [n_0; 2n_0]$; the tidal (with the mass) detections provide slightly tighter constraints than the radius.

With regards to the posterior’s ability to recover the injection EoS, for the PCP(BSK24) injection study, the posteriors tighten around the injection EoS across all density regimes for both low- and high-mass sources, as shown in Figs.~\ref{fig:post_1pct_small_bsk24} and ~\ref{fig:post_1pct_high_bsk24}. 
For the RG(SLY2) injection study, the simulated low-mass sources lead to posteriors that recover the injection EoS well across all density regimes, as shown in Fig.~\ref{fig:post_1pct_small_sly2}. When considering high-mass sources, as shown in Fig.~\ref{fig:post_1pct_high_sly2}, the posterior is tight but does not recover the injection EoS; moreover, as shown in Table~\ref{tab:effective_sample_size}, the effective posterior sample size is small. 
This is evident from Fig.~\ref{fig:obs_set_comparison/compare_mrl}: the macroscopic parameters of high-mass NSs governed by the EoS RG(SLY2) lie outside the 90\% contours of the prior; in other words, our EoS prior sets do not sufficiently explore part of the astrophysical parameter space, so the inference is statistically limited at high masses. For example, when considering NS mass constraints applied to the \expb set, EoSs with a non-zero likelihood account for only 45\% of the total number of EoSs in the prior when high-mass NSs are considered (high-mass NS can substantially reduce the prior), whereas they account for 98\% when low-mass NSs are considered.
A solution might be to employ semi-parametric EoSs with a high-density agnostic approach based on Gaussian processes, which offer a much wider prior at high density (relevant for high-mass NSs; see Fig.~2 in \citet{Ng2025}), or to use more sophisticated Monte Carlo methods to sample according to the likelihoods (e.g., Markov chain Monte Carlo or nested sampling).
For the GPPVA(NL3$\omega\rho$) injection study, the posteriors informed by simulated low masses do not recover the injection EoS well in the high-density regime; however, they performs well for ${n \lesssim 2n_0}$, as shown in Fig.~\ref{fig:post_1pct_small_nl3}.

Figure~\ref{fig:post_1pct} highlights that sufficiently wide NEP priors encompassing the NEPs of all injection EoSs do not guarantee the large statistical realization of EoSs that resemble the injection EoS in the prior. However, for all injection EoSs, the low-mass detections yield posteriors that recover the injection EoS well in the intermediate density regime $n\leq 2n_0$, which is the density regime of interest for NEP inference for this EoS set (\expb).

\subsubsection{Nuclear parameter inference} \label{sec:results/nuc_inference_Gaussian}

\begin{figure*}
    \centering
    \begin{subfigure}[b]{0.33\textwidth}
        \centering
        \resizebox{\hsize}{!}{\includegraphics{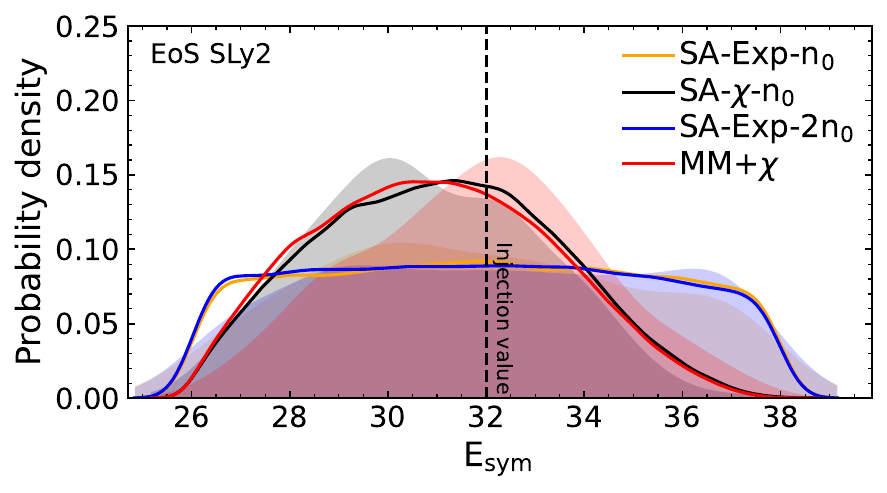}}
    \end{subfigure}
    \hfill
    \begin{subfigure}[b]{0.33\textwidth}
        \centering
        \resizebox{\hsize}{!}{\includegraphics{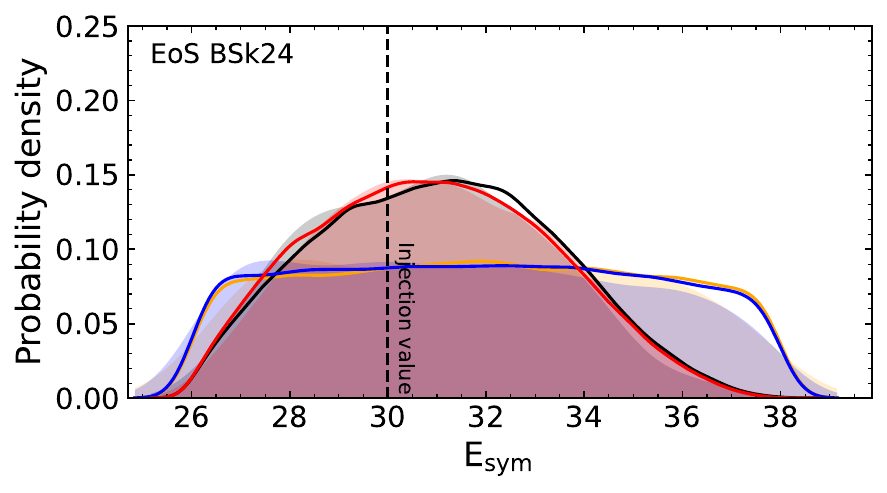}}
    \end{subfigure}
    \hfill
    \begin{subfigure}[b]{0.33\textwidth}
        \centering
        \resizebox{\hsize}{!}{\includegraphics{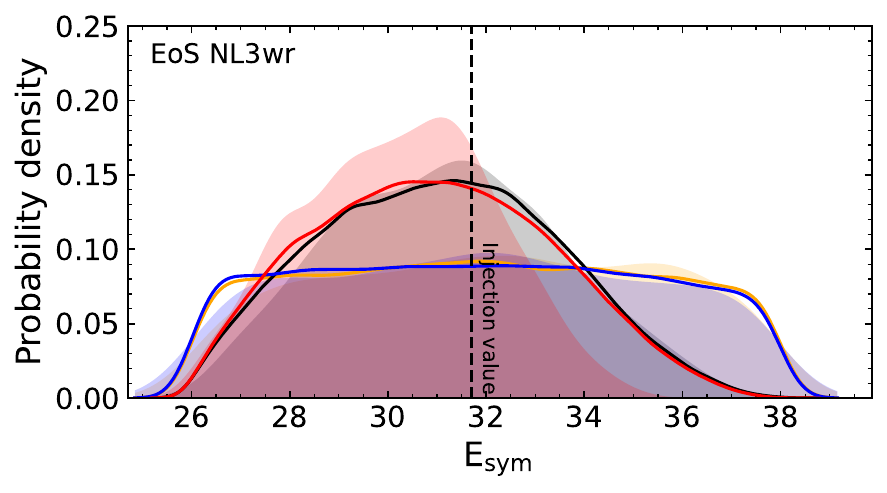}}
    \end{subfigure}
    \centering
    \begin{subfigure}[b]{0.33\textwidth}
        \centering
        \resizebox{\hsize}{!}{\includegraphics{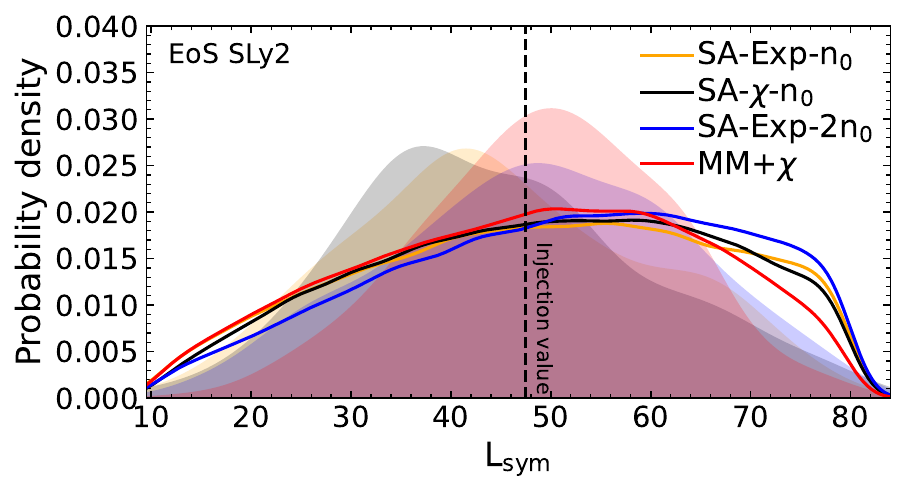}}
    \end{subfigure}
    \hfill
    \begin{subfigure}[b]{0.33\textwidth}
        \centering
        \resizebox{\hsize}{!}{\includegraphics{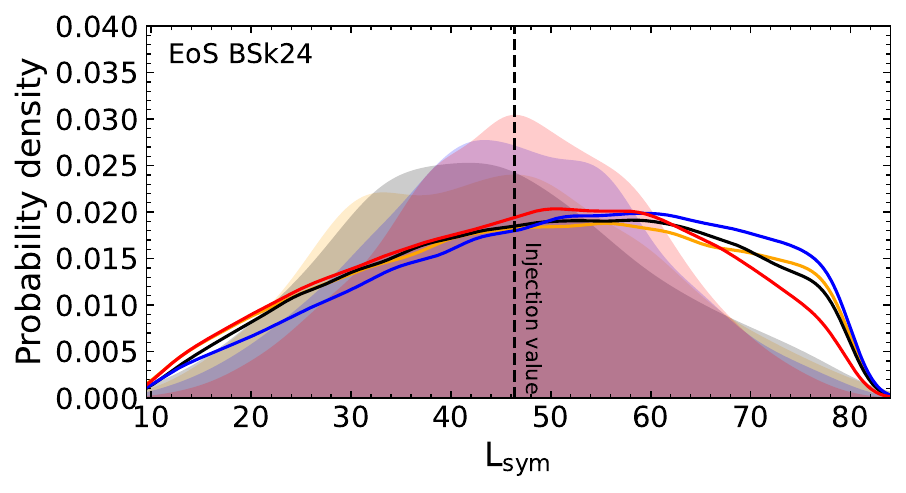}}
    \end{subfigure}
    \hfill
    \begin{subfigure}[b]{0.33\textwidth}
        \centering
        \resizebox{\hsize}{!}{\includegraphics{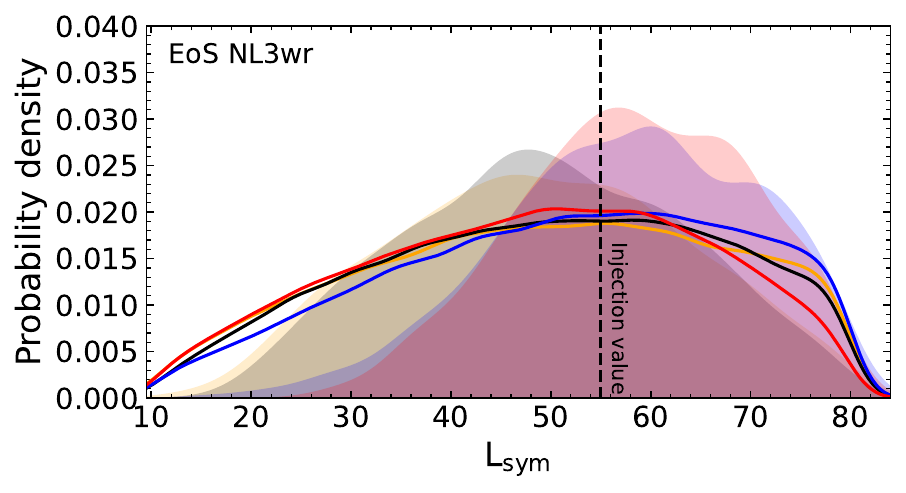}}
    \end{subfigure}
    \hfill
    \begin{subfigure}[b]{0.33\textwidth}
        \centering
        \resizebox{\hsize}{!}{\includegraphics{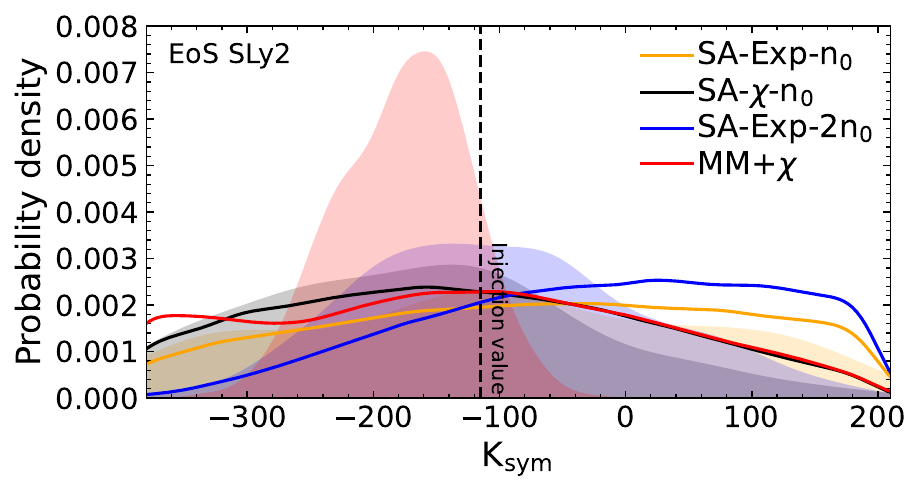}}
    \end{subfigure}
    \hfill
    \begin{subfigure}[b]{0.33\textwidth}
        \centering
        \resizebox{\hsize}{!}{\includegraphics{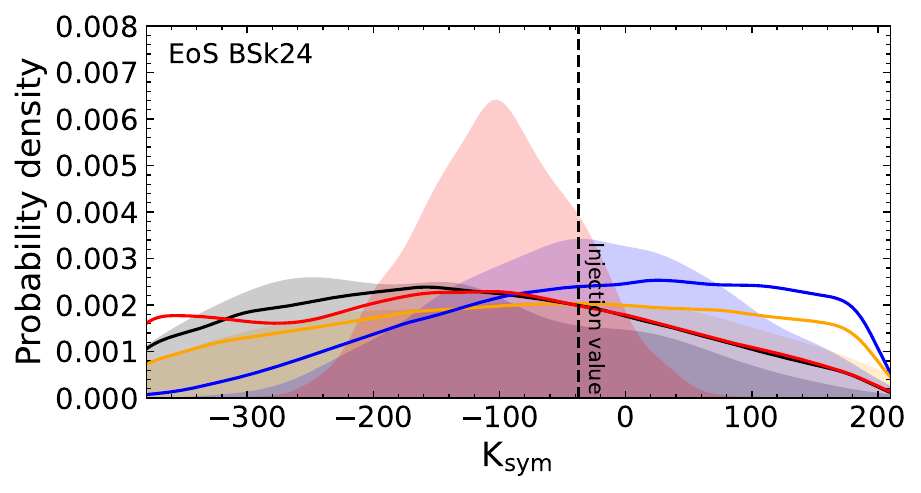}}
    \end{subfigure}
    \hfill
    \begin{subfigure}[b]{0.33\textwidth}
        \centering
        \resizebox{\hsize}{!}{\includegraphics{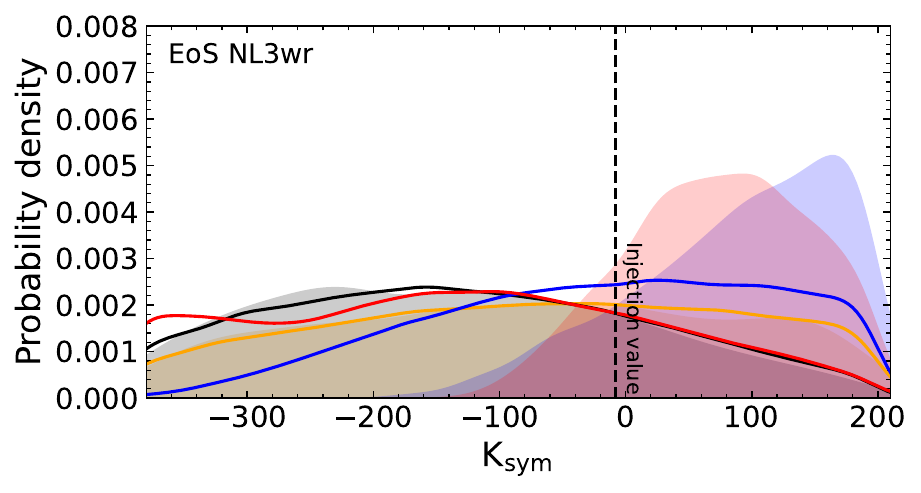}}
    \end{subfigure}
    \hfill
    \begin{subfigure}[b]{0.33\textwidth}
        \centering
        \resizebox{\hsize}{!}{\includegraphics{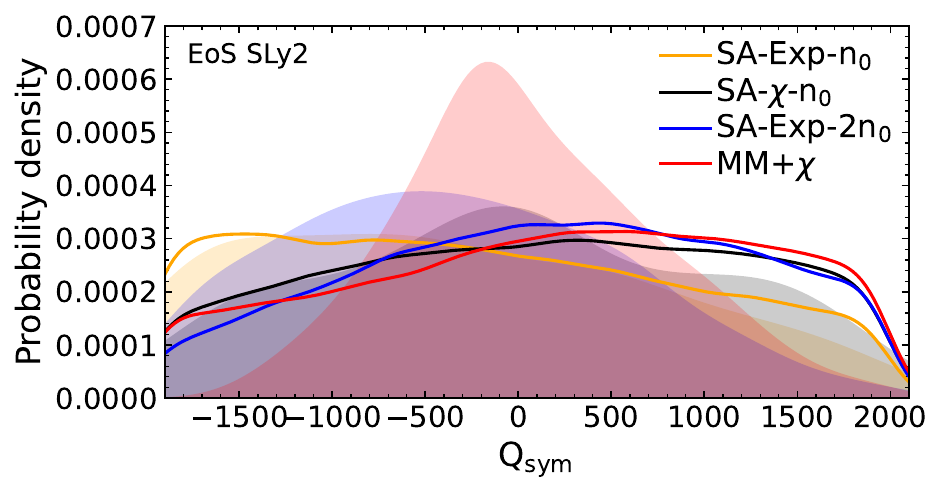}}
    \end{subfigure}
    \hfill
    \begin{subfigure}[b]{0.33\textwidth}
        \centering
        \resizebox{\hsize}{!}{\includegraphics{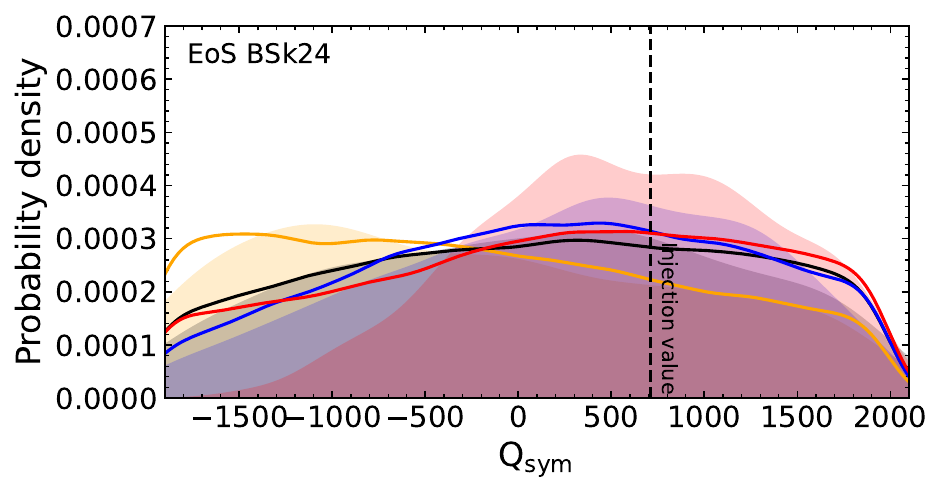}}
    \end{subfigure}
    \hfill
    \begin{subfigure}[b]{0.33\textwidth}
        \centering
        \resizebox{\hsize}{!}{\includegraphics{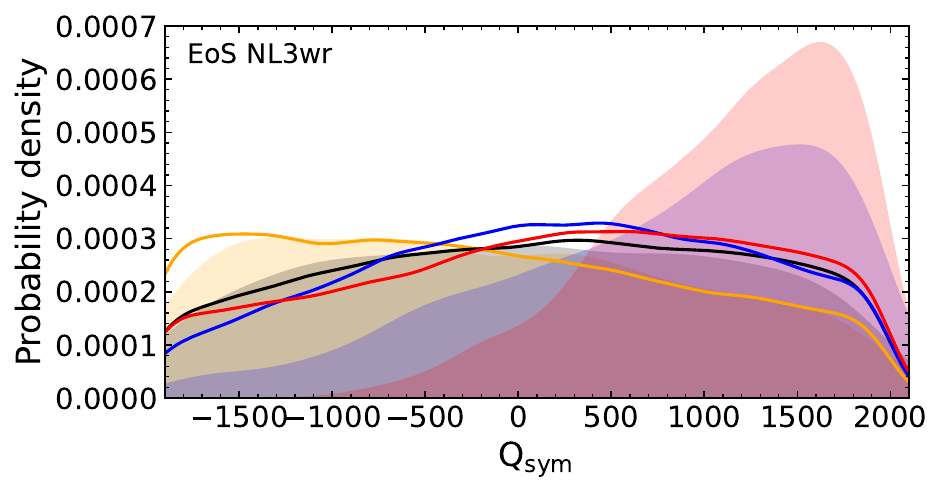}}
    \end{subfigure}
    \caption{Posterior distributions informed by $M-\Lambda$ simulations for \esym, \lsym, \ksym, and \qsym for low-mass NSs. We simulated the NSs with an uncorrelated bivariate Gaussian distribution with $\delta E=0.01$. Results are presented for \expa in orange, \expb in blue, and \mmset in red. For comparison, the prior distributions are shown as solid lines. The injection values of the NEPs for the EoSs RG(SLY2), PCP(BSK24), and GPPVA(NL3$\omega\rho$) are represented as vertical dashed black lines.}
    \label{fig:sym_bivariate_tides}
\end{figure*}

Only a few NEPs significantly correlate with the $\beta$-equilibrated pressure (see Fig.~\ref{fig:pearson_density}): \esat, \esym, \lsym, \ksym, and \qsym. The NEP \esat significantly correlates with the pressure in a very low-density regime: constraining this parameter requires observations capable of constraining the pressure in the outer crust of the star, which is not the case for our simulated data. Given the results presented in Sect.~\ref{sec:results/inference/eos} and Table~\ref{tab:effective_sample_size}, we restricted the NEP inference to that of \esym, \lsym, \ksym, and \qsym, based on low-mass $M-\Lambda$ simulated data. We present the prior distribution of the NEPs \esym, \lsym, \ksym, and \qsym, as well as the posterior distributions informed by low-mass NSs simulations, in Fig.~\ref{fig:sym_bivariate_tides} for $M-\Lambda$ detections.
We show additional results for $M-R$ detections in Fig.~\ref{fig:sym_bivariate_radius} of Appendix~\ref{sec:app/MR}, and we present a corner-plot representation of the posterior distribution of the isovector NEPs for sets \expb and \mmset, based on low-mass $M-\Lambda$ simulated data and the PCP(BSK24) injection, in Fig.~\ref{fig:corner} of Appendix~\ref{sec:app/corner}.

For \esym, the prior and astrophysically informed posterior distributions for the sets \expa and \expb coincide and are relatively flat (suggesting a uniform distribution similar to the NEP sampling)\footnote{We used a kernel density estimation to obtain the distributions presented in this section.}.
For the sets \chiset and \mmset, the prior and astrophysically informed posterior distributions are peaked. We conclude that the $\chi$-EFT data imposed in ${n \in [n_{\chi - \rm data}; n_0]}$ lead to already peaked priors and that astrophysical constraints do not significantly peak the distribution. 
The $\chi$-EFT data applied for $n \geq n_0$ do not help constrain this parameter, consistent with the conclusions drawn when discussing Fig.~\ref{fig:eos_set_comparison/compare}. Astrophysical information from low-mass simulations does not help recover the injection value of \esym.

For \lsym and \ksym, the astrophysical posteriors are more peaked than priors for all the sets; the prior distributions are not uniform, as the meta-model construction itself leads to NEP priors that depart from the NEP sampling.
For \lsym, the astrophysically informed posteriors for \expb and \mmset peak at the injection value for the injection EoS PCP(BSK24) (and to a lesser degree for RG(SLY2) and GPPVA(NL3$\omega\rho$)), suggesting that the density regime ${n\in [n_0; 2n_0]}$ plays a role in recovering \lsym, as indicated by Fig.~\ref{fig:pearson_density}.
For \ksym, the \mmset set presents a very peaked astrophysically informed posterior distribution that is systematically away from the injection value, suggesting that extending the meta-model EoS construction into the density regime $n \geq 2n_0$ pollutes the recovery of the injection value. This can be understood from the inability of the meta-model to reproduce well-known EoS models at high density (see discussion in~\citet{Margueron2018}), where higher orders of the Taylor expansion play an important role in determining the $\beta$-equilibrated pressure. 
The astrophysically informed posteriors for the \chiset set resemble its prior. For \expb in the RG(SLY2) and PCP(BSK24) injection studies, the posterior peaks at the injection value: we conclude that the density regime $n \in [n_0;2n_0]$ plays an important role in constraining \ksym, and that the semi-agnostic \expb set is capable of recovering the injection value of the NEP, unlike \mmset. However, the injection study with EoS GPPVA(NL3$\omega\rho$) yields a peaked posterior distribution far from the injection value, although Fig.~\ref{fig:post_1pct} shows that the $\beta$-equilibrated pressure from low-mass simulations is recovered well within the 90\% contours for $n \in [n_0; 2n_0]$. 
This can be explained by the inability of the \expb prior set to sufficiently explore the NEP parameter space. 
The injection-value point of GPPVA(NL3$\omega\rho$) in the \esym-\ksym parameter space is located close to a region where the density lines present holes (or dips) in probability density. This suggests that, although the boundaries of the uniform distributions sampled for the NEPs (see Table~\ref{tab:nuc_param}) encompass the injection values of the NEPs, they do not guarantee that the set of NEPs after the meta-model construction is uniformly distributed in the large dimensionality of the meta-model.

For \qsym, the \mmset set presents the most peaked astrophysically informed distributions. We note that the injection value for \qsym is only available for the EoS PCP(BSK24), and therefore conclusions on the recovery of the parameter over a large stiffness range were not possible in this study. However, the figure confirms that the density regime $n\geq 2n_0$ strongly correlates with this parameter, which is a high-order parameter in the meta-model Taylor expansion.

\section{Discussion}
\label{sec:discussions}

The semi-agnostic EoS construction presented in this paper can be amended to improve the statistical realization of NEP and EoS priors. In the future, our goal is to implement several improvements to \texttt{CUTER}.
For the high-density EoS, piecewise polytropes offer less flexibility than so-called nonparametric approaches. 
The high-density flexibility of the EoS prior can thus be improved by using, for example, Gaussian processes as a high-density extension of our semi-agnostic approach (see \citet{Ng2025}). This will improve the recovery of the radius and tidal deformability in the injection study for high-mass detections. Additionally, Gaussian-process high-density extensions that are causal by construction would allow us to impose pQCD constraints and better evaluate their impact on nuclear parameter inference from high-mass neutron star detections.

Moreover, in this study, we used precomputed sets containing 50000 EoSs each. Increasing the number of EoSs in the sets is a simple way to improve the smoothness of the astrophysical posterior EoS contours. However, to obtain  posteriors with a statistically significant number of EoS and NEP multidimensional realization, it may be necessary to generate EoSs concurrently with Bayesian inference, that is,  according to the astrophysical detection likelihood using a smart sampler (also called sampling on the fly). We note that this approach would be incompatible with the current implementation of nonparametric high-density extensions. 
    
\citet{Lim2024} discuss an improved version of the meta-model construction.
Studies such as \citet{Mondal2023, Klausner2025} also suggest resampling the higher order parameters to improve the flexibility of the meta-model at high density and to ensure that the low-density behavior of the EoS does not impose spurious correlations on the high-density regime.
Amending the meta-model construction in our semi-agnostic models could improve the inference of NEPs, especially for high values of $n_{\rm match}$.

We could also improve the injection study by employing more sophisticated simulations of astrophysical data. Accounting for correlations between the astrophysical parameters in GW and X-ray pulse-profile modeling, as well as the sensitivities of detectors such as the Einstein Telescope, Cosmic Explorer, or ATHENA, would clarify the prospects for NEP inference in the context of the next generation of telescopes.

\section{Conclusions} 
\label{sec:conclusion}

We have explored the capability of semi-agnostic EoS priors to recover NEPs through injection studies based on three nucleonic EoSs and simple simulations of mass, radius, and tidal-deformability detections.
We show that semi-agnostic priors with piecewise polytropes attached at $n = n_0, 2n_0$ offer much greater EoS flexibility than the full meta-model, in agreement also with previous findings (see e.g., \citet{Tews2018} who compare speed-of-sound models with a meta-model approach).
Our injection studies based on nucleonic EoSs RG(SLY2), PCP(BSK24), and GPPVA(NL3$\omega\rho$), show that astrophysical detections constrain the $\beta$-equilibrated pressure in different density regimes for low- or high-mass neutron-star detections.

We find that not all NEPs correlate significantly with the $\beta$-equilibrated pressure in density regimes relevant for astrophysical observations of the mass, radius, and tidal deformability of NSs. We conclude that only a limited number of NEPs can be constrained by these astrophysical detections. 
We also show that semi-agnostic models can recover the injected \ksym more accurately than the full meta-model, as the latter can produce posterior distributions that are strongly peaked away from the injection value of the NEP. 
Finally, we discuss how to improve the EoS construction to ensure sufficient sampling of the NEP distribution and to realize the astrophysical parameters.

\begin{acknowledgements}
The authors thanks the developers of
\texttt{CUTER}. L.S acknowledges the financial support of the National Science Foundation Grant No. PHY 21-16686. J.R. acknowledges support from the National Science Foundation Grant No. PHY 2409736. This work is supported by Deutsches Elektronen Synchrotron and Deutsches Centrum f\"ur Astrophysik. This work has been also partially supported by the IN2P3 Master Project MAC, and the CNRS International Research Project (IRP) `Origine des \'el\'ements lourds dans l’univers: Astres Compacts et Nucl\'eosynth\`ese (ACNu)'.
\end{acknowledgements}

\bibliographystyle{aa}
\bibliography{biblio}
\newpage

\begin{appendix}

\section{Mass-radius and Mass-tidal deformability for the different sets} \label{sec:app/mr-ml}

\begin{figure}[H]
    \centering
    \resizebox{\hsize}{!}{\includegraphics{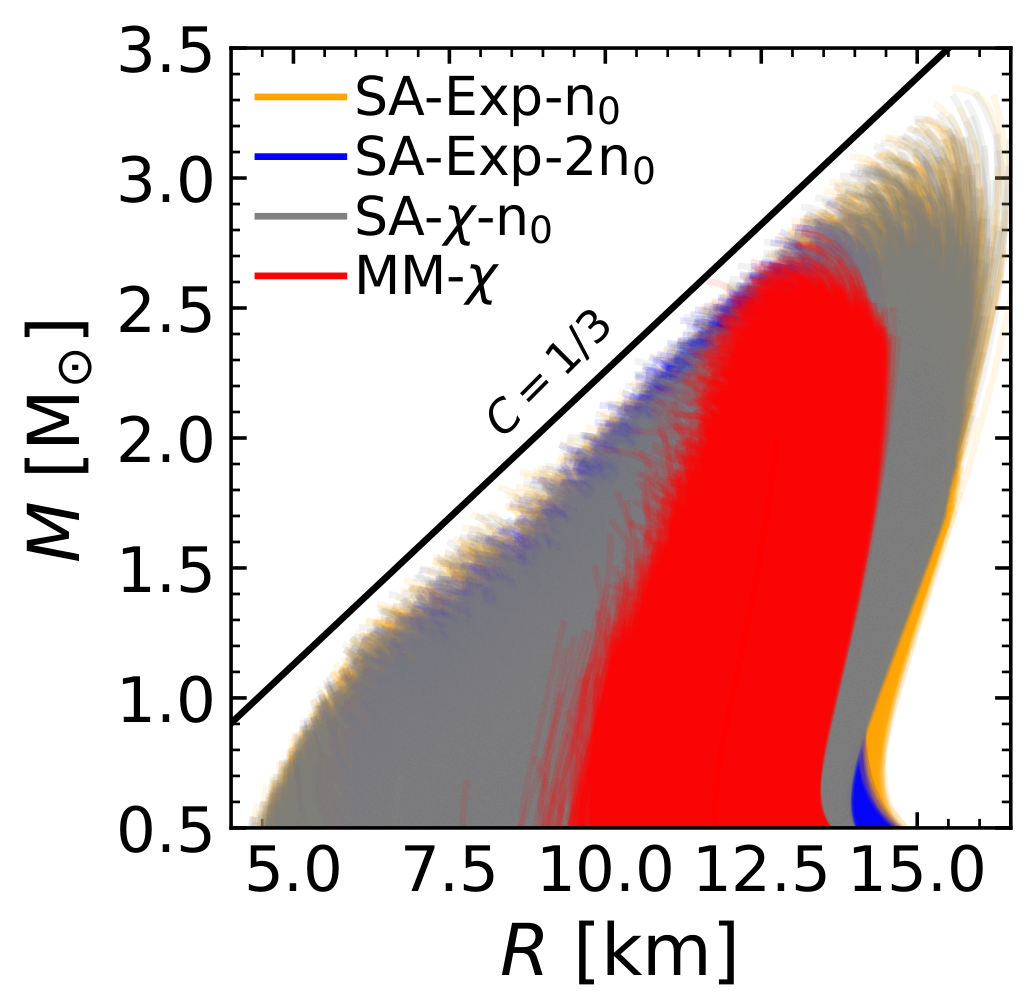}}
    \caption{Mass-radius relation for all EoSs of the \expa, \expb, \chiset and \mmset sets. The compactness limit $C=1/3$.}
    \label{fig:full_MR-all}
\end{figure}

We present the mass-radius relations for every EoS of the four different sets in Fig.~\ref{fig:full_MR-all}. All four different sets respect the compactness limit discussed in Ref.~\cite{Rezzolla2026}.

\begin{table*}[]
    \centering
    \caption{Effective sample size of the $M-R$ posterior for each injection EoS and EoS set presented in this paper. }
    \begin{tabular}{|c|c|c|c|c|}
        \hline
        & \expa & \expb & \chiset & \mmset \\ \hline 
        & \multicolumn{4}{|c|}{Injection EoS RG(SLY2)}\\ \hline \hline 
        Low-masses & [2035 - 2502]  & [4540 - 5498] & [2028 - 2524] & [3291 - 4257]\\ \hline 
        High-masses & [117 - 428] & [110 - 647] & [108 - 417] & [1 - 86] \\ \hline
        & \multicolumn{4}{|c|}{Injection EoS PCP(BSK24)}\\ \hline \hline 
        Low-masses & [3328 - 4170]& [9364 - 9721] & [3468 - 4294] & [7361 - 7743] \\ \hline
        High-masses & [680 - 1175] & [1579 - 3162] & [737 - 1214] & [1228 - 4316]  \\ \hline
        & \multicolumn{4}{|c|}{Injection EoS GPPVA(NL3$\omega\rho$)}\\ \hline \hline 
        Low-masses & [8952 - 9390] & [7021 - 7809] & [9472 - 10213] & [4087 - 4127]\\ \hline
        High-masses & [3039 - 4329] & [2468 - 3736] & [3077 - 4415] & [4037 - 4042]\\ \hline
    \end{tabular}
    \tablefoot{Results are presented as intervals (pertaining to the 10 sources considered) for the low- and high-mass NSs simulated for our study.}
    \label{tab:effective_sample_size}
\end{table*}

\section{Effective posterior sample size}\label{sec:app/size}

In Table.~\ref{tab:effective_sample_size}, we present the effective sample size of the posteriors for different EoS sets informed by mass-radius simulated data; note that the orders of magnitudes are similar if mass-tidal deformability detections are considered. We find that the NS mass is a key factor that can decimate the effective sample sizes of the posterior. The latter ones are smaller when considering EoS sets with low values of $n_{\rm match}$: sets with $n_{\rm match} = n_0$ present larger priors than sets with $n_{\rm match} = 2n_0$, making the pressure-density space more dense for the latter, and resulting in generally larger effective sample for a given simulated data. This is also confirmed by the effective sample size of \mmset. Finally, when considering high-mass simulated data and the injection EoS RG(SLY2), the effective sample size of the posteriors is small due to the inability of the EoS constructions to reproduce the corresponding NS astrophysical parameters.

\section{$M-R$ informed posterior distribution of NEPs}\label{sec:app/MR} 

We find that the astrophysically informed posterior distribution are quite similar in the case of $M-R$ detections and $M-\Lambda$ detections. We find slightly more peaked distribution for \esym and and \lsym when considering radius detection, and slightly less peaked distribution for \ksym informed by the radius than by the tidal deformability. This can be understood by the sensitivity of the radius on lower density regimes than tidal deformability, and by the fact that \ksym appears at a higher order in the meta-model Taylor expansion with respect to \esym and \lsym. 
\begin{figure*}
    \centering
    \begin{subfigure}[b]{0.33\textwidth}
        \centering
        \resizebox{\hsize}{!}{\includegraphics{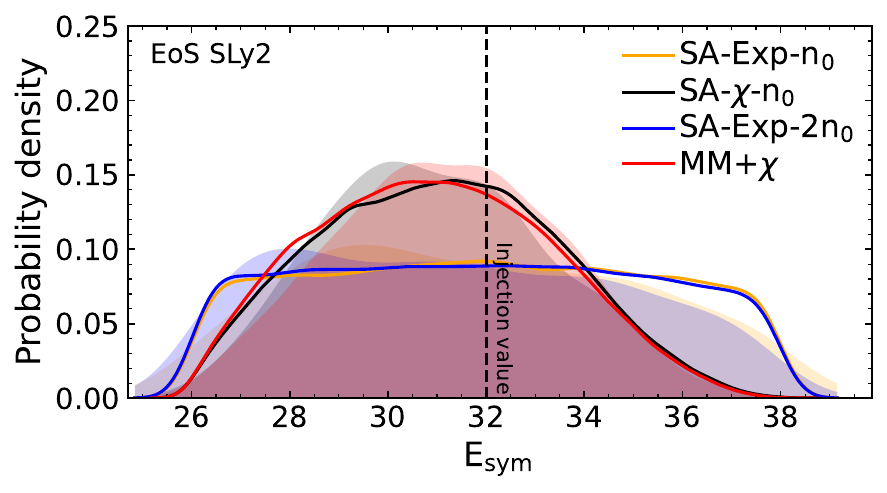}}
    \end{subfigure}
    \hfill
    \begin{subfigure}[b]{0.33\textwidth}
        \centering
        \resizebox{\hsize}{!}{\includegraphics{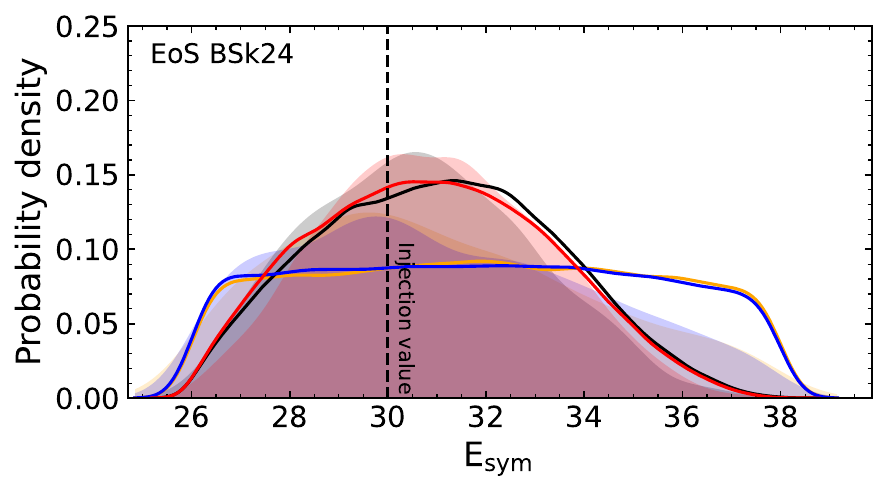}}
    \end{subfigure}
    \hfill
    \begin{subfigure}[b]{0.33\textwidth}
        \centering
        \resizebox{\hsize}{!}{\includegraphics{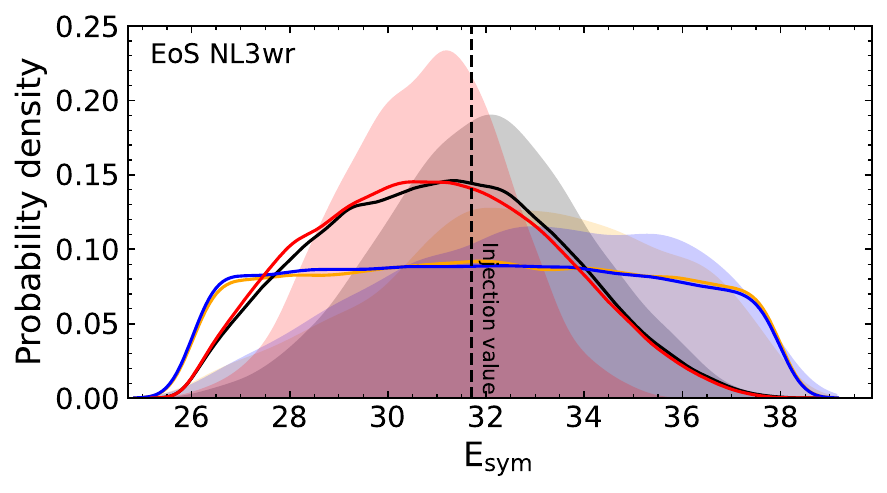}}
    \end{subfigure}
    \centering
    \begin{subfigure}[b]{0.33\textwidth}
        \centering
        \resizebox{\hsize}{!}{\includegraphics{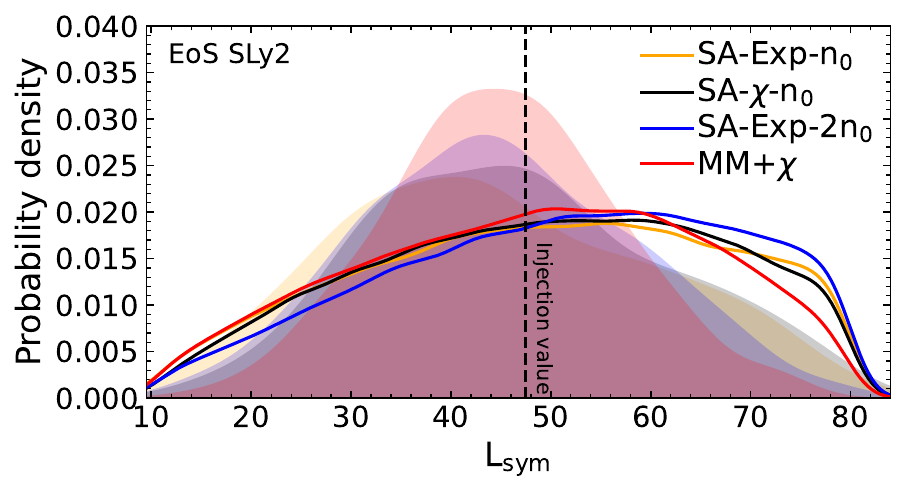}}
    \end{subfigure}
    \hfill
    \begin{subfigure}[b]{0.33\textwidth}
        \centering
        \resizebox{\hsize}{!}{\includegraphics{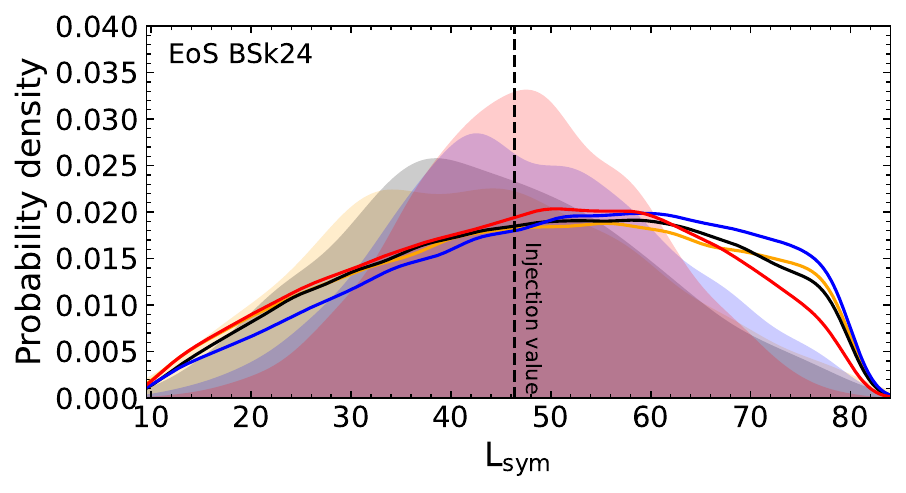}}
    \end{subfigure}
    \hfill
    \begin{subfigure}[b]{0.33\textwidth}
        \centering
        \resizebox{\hsize}{!}{\includegraphics{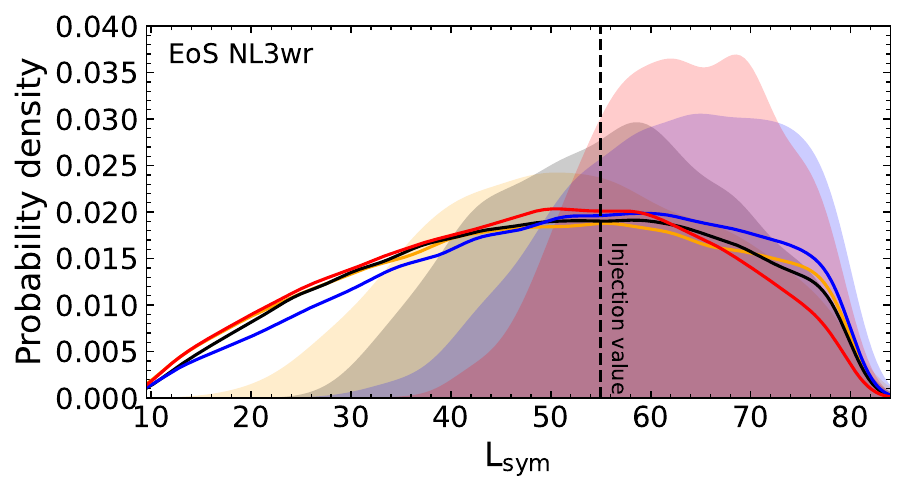}}
    \end{subfigure}
    \hfill
    \begin{subfigure}[b]{0.33\textwidth}
        \centering
        \resizebox{\hsize}{!}{\includegraphics{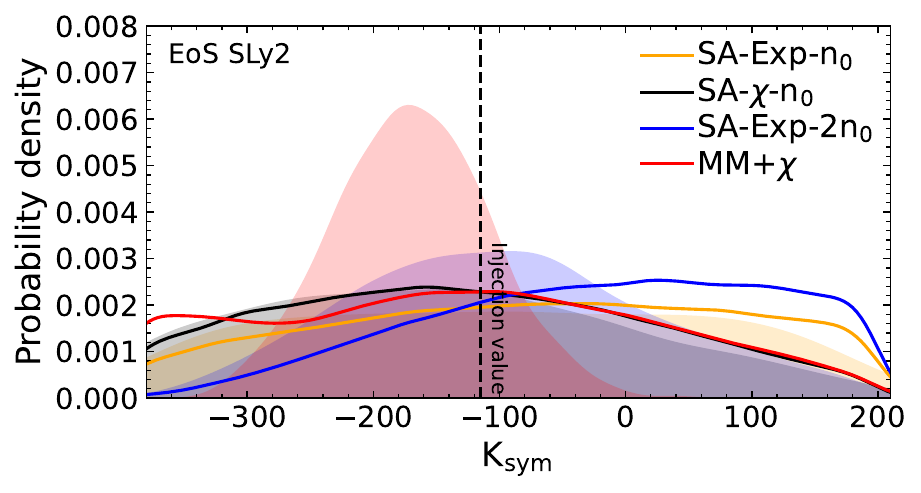}}
    \end{subfigure}
    \hfill
    \begin{subfigure}[b]{0.33\textwidth}
        \centering
        \resizebox{\hsize}{!}{\includegraphics{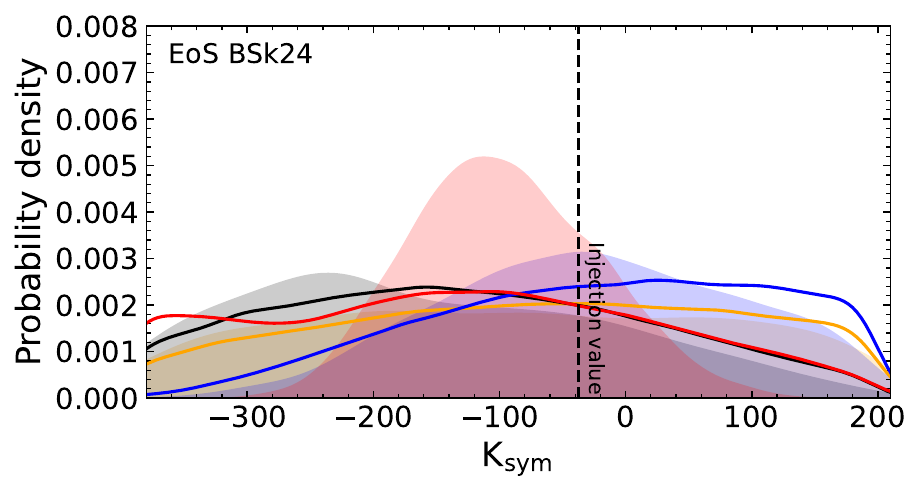}}
    \end{subfigure}
    \hfill
    \begin{subfigure}[b]{0.33\textwidth}
        \centering
        \resizebox{\hsize}{!}{\includegraphics{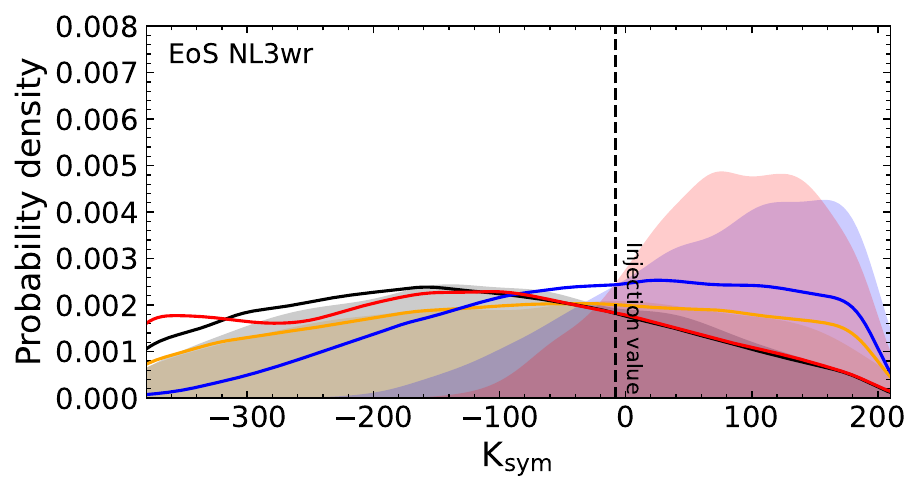}}
    \end{subfigure}
    \hfill
    \begin{subfigure}[b]{0.33\textwidth}
        \centering
        \resizebox{\hsize}{!}{\includegraphics{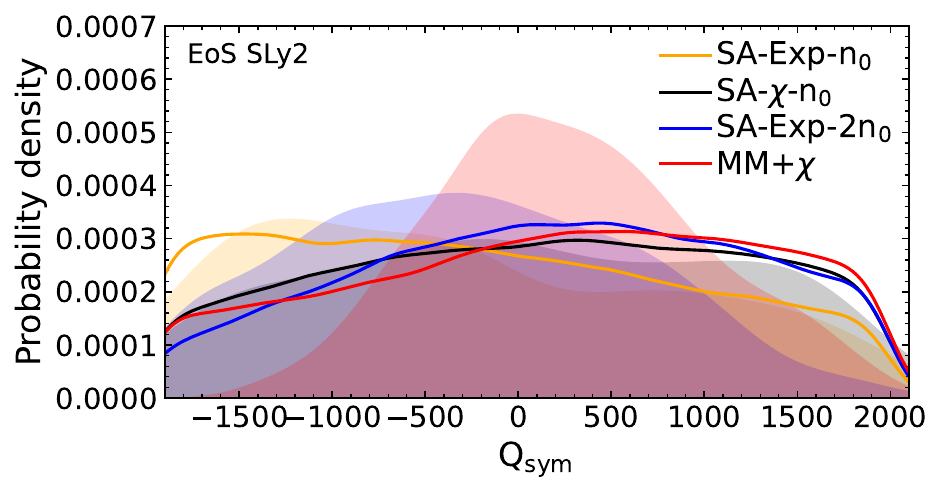}}
    \end{subfigure}
    \hfill
    \begin{subfigure}[b]{0.33\textwidth}
        \centering
        \resizebox{\hsize}{!}{\includegraphics{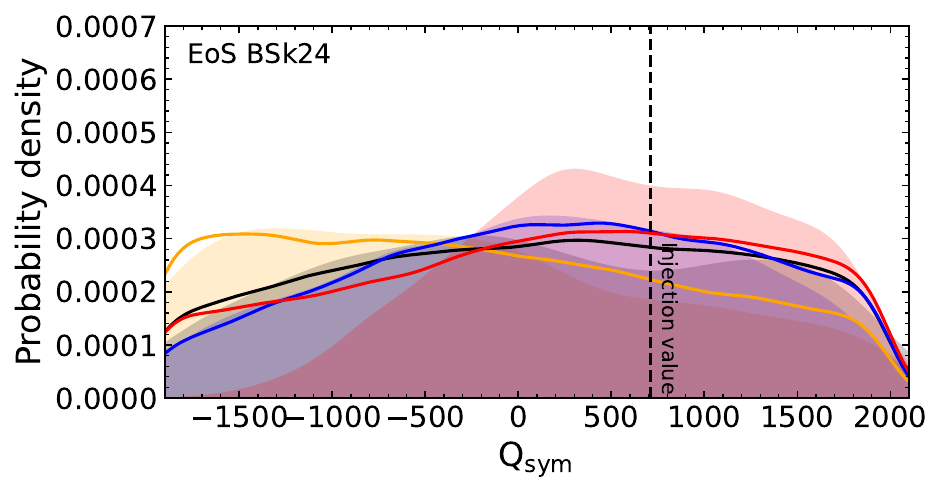}}
    \end{subfigure}
    \hfill
    \begin{subfigure}[b]{0.33\textwidth}
        \centering
        \resizebox{\hsize}{!}{\includegraphics{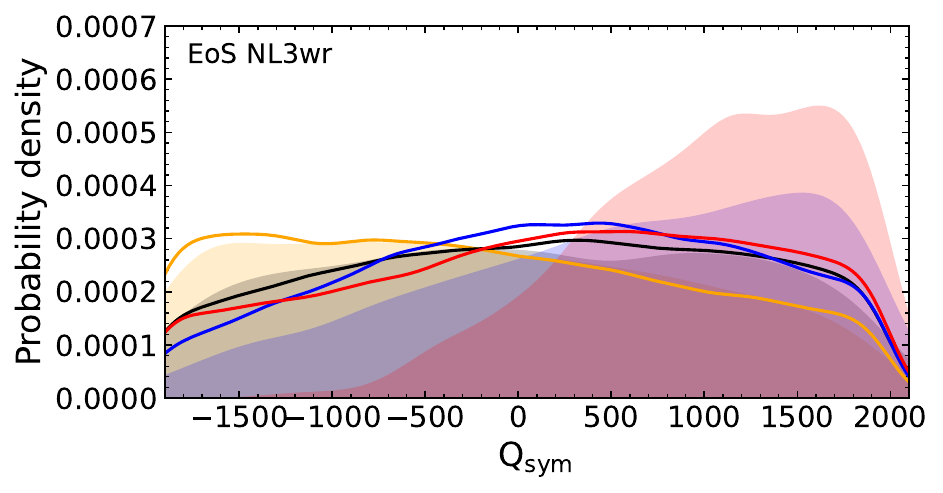}}
    \end{subfigure}
    \caption{$M-R$ informed posterior distribution for \esym, \lsym, \ksym and \qsym for low-mass NSs simulated with an uncorrelated bivariate Gaussian distribution with $\delta E=0.01$. Results are presented for \expa in orange, \expb in blue and \mmset in red. For comparison, the prior distributions are shown with a plain line. The injection values of the NEPs for EoSs RG(SLY2), PCP(BSK24) and GPPVA(NL3$\omega\rho$) are represented as a vertical black dashed line.}
    \label{fig:sym_bivariate_radius}
\end{figure*}

\section{$M-L$ informed posterior distribution comparison between \expb and \mmset}\label{sec:app/corner}

\begin{figure*}
    \sidecaption
    \includegraphics[width=12cm]{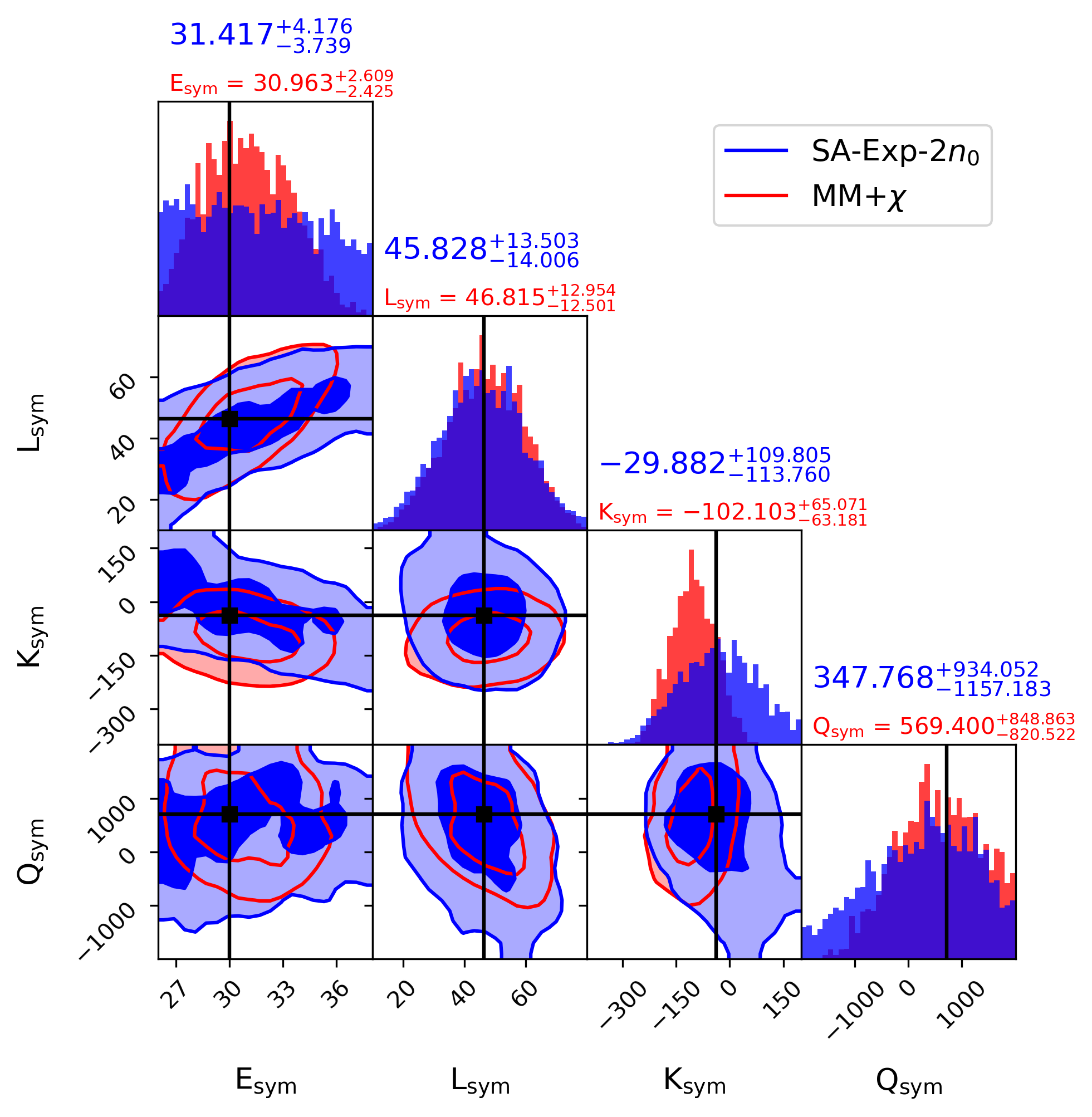}
    \caption{Isovector NEP posterior distributions informed by $M-\Lambda$ detections of low-mass neutron stars simulated with PCP(BSK24), represented with 1-$\sigma$ and 2-$\sigma$ contours. Results for \expb are presented in blue, and for \mmset in red. Injection value for PCP(BSK24) are represented in black.}
    \label{fig:corner}
\end{figure*}

In Fig.~\ref{fig:corner}, we present the corner-plot of the posterior distribution of the isovector NEPs informed by low-mass neutron stars simulated with PCP(BSK24). 
We compare results for the \mmset and \expb prior sets. 
This figure confirms the findings presented in Sect.~\ref{sec:results/nuc_inference_Gaussian}, namely, (i) a larger posterior distribution for the NEPs for the \expb set (except for $L_{\rm sym}$, for which the width of the distribution is comparable in the two sets; see Fig.~\ref{fig:sym_bivariate_tides}), and (ii) a better recovery of the parameters for the semi-agnostic (\expb) set with respect to the \mmset set.
Moreover, we observe, particularly for the \mmset set, a (slight) correlation between $L_{\rm sym}$ and $E_{\rm sym}$, between $K_{\rm sym}$ and $L_{\rm sym}$, and between $Q_{\rm sym}$ and $K_{\rm sym}$. These findings are in agreement with those of \citet{Dinh2021Uni} (see their Fig.~12 and related discussion).
We also notice, for the same set, a slight correlation between $K_{\rm sym}$ and $E_{\rm sym}$.

\end{appendix}

\end{document}